\documentclass[conference,compsoc]{IEEEtran}
\IEEEoverridecommandlockouts
\usepackage{cite}
\usepackage{makecell}
\usepackage{booktabs}
\usepackage{amsthm} 
\usepackage[normalem]{ulem}
\newtheorem{theorem}{Theorem}
\usepackage{multirow}
\usepackage{url}
\usepackage{amssymb,amsfonts}
\usepackage{amsmath}
\usepackage{algorithm}
\usepackage{algpseudocode}
\usepackage{graphicx}
\usepackage{textcomp}
\usepackage{xcolor}
\usepackage{fancyhdr}
\usepackage{subfigure}
\usepackage[nonumberlist]{glossaries}
\makeglossaries
\setglossarystyle{list}

\usepackage[left=1.62cm,right=1.62cm,top=1.9cm, bottom=2.67cm]{geometry}
\def\BibTeX{{\rm B\kern-.05em{\sc i\kern-.025em b}\kern-.08em
    T\kern-.1667em\lower.7ex\hbox{E}\kern-.125emX}}

\newboolean{showcomments}
\setboolean{showcomments}{true}
\ifthenelse{\boolean{showcomments}}
{
}

\newcommand*{\affmark}[1][*]{\textsuperscript{#1}}

\newacronym{RL}{RL}{Reinforcement Learning}
\newacronym{LLM}{LLM}{Large Language Model}
\newacronym{GenAI}{GenAI}{Generative AI}
\newacronym{LoRA}{LoRA}{Low Rank Adaptation}
\newacronym{QLoRA}{QLoRA}{Quantized Low Rank Adapter}
\newacronym{DT}{DT}{Decision Transformer}
\newacronym{HDTGA}{HDTGA}{Hierarchical Decision Transformer with Goal Awareness}
\newacronym{NLP}{NLP}{Natural Language Processing}
\newacronym{HRL}{HRL}{Hierarchical RL}
\newacronym{LSTF}{LSTF}{Long Sequence Time Series Forecasting}
\newacronym{RAN}{RAN}{Radio Access Network}
\newacronym{CDL}{CDL}{Clustered Delay Line}
\newacronym{LOS}{LOS}{Line-of-Sight}
\newacronym{NLOS}{NLOS}{Non-Line-of-Sight}
\newacronym{UE}{UE}{User Equipment}
\newacronym{BS}{BS}{Base Station}
\newacronym{QoS}{QoS}{Quality of Service}
\newacronym{IRCs}{IRCs}{Intelligent RAN Controllers}
\newacronym{DQN}{DQN}{Deep-Q-Network}
\newacronym{h-DQN}{h-DQN}{hierarchical Deep-Q-Network}
\newacronym{MDP}{MDP}{Markov Decision Process}
\newacronym{TTI}{TTI}{Transmission Time Interval}
\newacronym{KPI}{KPI}{Key Performance Indicator}
\newacronym{GPT}{GPT}{Generative Pretrained Transformer}
\newacronym{DOT}{DOT}{Decoder Only Transformer}
\newacronym{LSTM}{LSTM}{Long Short-Term Memory}
\newacronym{RAG}{RAG}{Retrieval Augmented Generation}
\newacronym{SLA}{SLA}{Service Level Agreement} 
\newacronym{Open RAN}{Open RAN}{Open Radio Access Network}
\newacronym{OFDM}{OFDM}{Orthogonal Frequency Division Multiplexing}
\newacronym{DRL}{DRL}{Deep RL}
\newacronym{SINR}{SINR}{Signal-to-Interference-plus-Noise Ratio}
\newacronym{GPU}{GPU}{Graphics Processing Unit}
\newacronym{CPU}{CPU}{Central Processing Unit}
\newacronym{PDCCH}{PDCCH}{Physical Downlink Control Channel}
\newacronym{RAT}{RAT}{Radio Access Technology}

\begin{document}

%\title{Generative AI on the Air: An End-to-End Framework for Intent-Based Automation in 5G and Beyond Networks}
%\title{An End-to-End Framework for Generative AI-based Intent-driven Network Management in 6G}
\title{Harnessing the Power of LLMs, Informers and Decision Transformers for Intent-driven RAN Management in 6G}
\author{\IEEEauthorblockN{Md~Arafat~Habib\affmark[1] \IEEEmembership{Graduate Student Member,~IEEE}, Pedro~Enrique~Iturria-Rivera\affmark[2], Yigit Ozcan\affmark[2], \\Medhat Elsayed\affmark[2], Majid Bavand\affmark[2], Raimundas Gaigalas\affmark[3], and  Melike Erol-Kantarci\affmark[1], \IEEEmembership{Fellow,~IEEE}}
\IEEEauthorblockA{\affmark[1]\textit{School of Electrical Engineering and Computer Science, University of Ottawa, Ottawa, Canada}}  \affmark[2]\textit{Ericsson Inc., Ottawa, Canada}, \affmark[3]\textit{Ericsson Inc., Stockholm, Sweden}\\
Emails:\{mhabi050,melike.erolkantarci\}@uottawa.ca, \\\{pedro.iturria.rivera, yigit.ozcan, medhat.elsayed, majid.bavand, raimundas.gaigalas\}@ericsson.com \vspace{-1em}}

\maketitle 
\thispagestyle{fancy}   
\fancyhead{}                
\lhead{Submitted for possible publication to IEEE. Paper currently under review. The contents of this paper may change at any time without notice.}
\cfoot{}
\renewcommand{\headrulewidth}{0pt} 

\begin{abstract}     

Intent-driven network management is critical for managing the complexity of 5G and 6G networks. It enables adaptive, on-demand management of the network based on the objectives of the network operators. In this paper, we propose an innovative three-step framework for intent-driven network management based on Generative AI (GenAI) algorithms. First, we fine-tune a Large Language Model (LLM) on a custom dataset using a Quantized Low-Rank Adapter (QLoRA) to enable memory-efficient intent processing within limited computational resources. A Retrieval Augmented Generation (RAG) module is included to support dynamic decision-making. Second, we utilize a transformer architecture for time series forecasting to predict key parameters, such as power consumption, traffic load, and packet drop rate, to facilitate intent validation proactively. Lastly, we introduce a Hierarchical Decision Transformer with Goal Awareness (HDTGA) to optimize the selection and orchestration of network applications and hence, optimize the network. Our intent guidance and processing approach improves BERTScore by $6\%$, and semantic similarity score by $9\%$ compared to the base LLM model. Again, the proposed predictive intent validation approach can successfully rule out the performance-degrading intents with an average of $88\%$ accuracy. Finally, compared to the baselines, the proposed HDTGA algorithm increases throughput at least by $19.3\%$, reduces delay by $48.5\%$, and boosts energy efficiency by $54.9\%$.
\end{abstract}


\section{Introduction}

An intent represents a set of expectations, including requirements, objectives, and limitations for a specific service or network management workflow \cite{1}. It is designed to be human-understandable while being precisely interpretable by machines. An intent specifies what needs to be achieved, focusing on the desired outcomes. Without detailing how to accomplish those outcomes, it provides target metrics with no possible elaboration on the implementation steps. Advancements in \gls{NLP} can enable network operators to express intents more effectively for intent-driven network management \cite{3}. This will provide simplification in managing vast and complex network infrastructures by focusing on what the network should achieve rather than how to configure it manually. Manual configurations are reduced in intent-driven network management, leading to fewer errors and increased operational efficiency. Furthermore, this kind of management supports rapid configuration changes and service deployment without manual intervention \cite{43}.

In 5G and envisioned 6G networks, implementing intent-driven network management can be challenging. An intent that can truly provide significant and relevant performance improvements, depends on network conditions, user demands, and regulations. As a result, a contextual understanding of intents is essential for accurate processing. Prior rule-driven, static, and manual policy-oriented \gls{SLA}-based approaches may fail to understand such contexts. On the contrary, a \gls{LLM}, with its transformer-based architecture, is well-suited for capturing the context needed and interpreting the underlying meaning of intents \cite{2}. Approaches like \cite{3,4,5} have put significant efforts into intent processing via \gls{LLM}s. However, they lack resource efficiency and adaptability to dynamic network scenarios. Moreover, these methods are not tailored to facilitate customized intent processing; a customization that can result in enhanced task-specific performance. 

Processing intents expressed in natural language is necessary to generate policies that can optimize network performance. However, allowing any intent to directly affect network configurations can lead to performance and stability issues. For example, an intent like ``Increase energy efficiency by $30\%$” during peak hours could degrade performance as there is high traffic load at the same time. This scenario highlights the need for a robust intent validation process that considers current network conditions to ensure smooth user experience without compromising \gls{QoS}. Previous works have considered rule-based approaches for intent validation. For instance, in \cite{2}, the rule-based validation approach ensures policies that adhere to predefined formats. However, it lacks adaptability to anticipated network changes. 

Emerging architectures such as \gls{Open RAN} \cite{6} have provided possibilities for third-party vendors to develop network-optimizing applications. Examples of such applications include traffic steering, beamforming, power allocation, and so on. However, managing the initiation and orchestration of these applications is a great challenge. The applications may conflict with each other if initiated altogether. An example of such a scenario may occur if applications performing traffic steering and cell sleeping are initiated together since they may have conflicting network objectives. On the other hand, power allocation with beamforming applications can significantly boost performance. To optimally initiate and orchestrate applications, past works have used advanced \gls{RL} algorithms. Habib et al. in their works \cite {7,8} have used \gls{HRL} to manage and orchestrate these applications. However, policies trained for a specific task using these methods may not perform well on similar tasks with slight changes. They react to current states without utilizing the sequence of past states and actions.

Based on the discussion so far, we summarize the following limitations of the existing intent processing, validation, and orchestration methodologies. First, intent processing approaches using \gls{LLM}s lack resource efficiency, adaptability to dynamic network scenarios, and customization for task-specific actions. Second, rule-based intent validation methods ensure policies adhere to predefined formats but fail to adapt to anticipated network changes. Third, \gls{RL} methods, including \gls{HRL}, used for application orchestration, are limited by their inability to generalize to similar tasks with slight changes. Also, \gls{RL} algorithms heavily rely on current states, without utilizing past state-action sequences. This restricts their decision-making capabilities.

Research in the domain of AI-enabled wireless networks is experiencing a paradigm shift towards \gls{GenAI} after the introduction of the transformer architecture and its demonstrated effectiveness \cite{35}. To this end, we propose a three-step solution based on \gls{GenAI} to address the problems discussed so far associated with intent processing, validation, and network application orchestration. The proposed methodology features an end-to-end \gls{GenAI} solution, providing complete intent-driven network automation for modern-day mobile communication systems.  

First, we build a custom dataset that has queries and prompts with their responses to fine-tune an LLM using a memory and resource-efficient technique called \gls{QLoRA} \cite{9}. Fine-tuning the LLM with our dataset ensures that it is specifically adapted to our queries and intents. To keep up with the dynamic network scenarios, we further use a \gls{RAG} module \cite{10} for taking real-time decisions with the help of an updated knowledge base. 

Second, a novel transformer architecture named Informer \cite{11} is used to perform \gls{LSTF}. It predicts three crucial parameters in future time slots: power consumption, traffic load, and packet drop percentage. Intents are validated based on these values. It significantly helps to avoid any intent that can have a negative impact on the system's performance.

Lastly, a new machine learning paradigm named \gls{DT} is employed to take control decisions regarding which network applications to initiate and how to orchestrate them \cite{12}. In particular, we propose a \gls{HDTGA}. Instead of relying on fixed reward sequences (as in traditional \gls{DT}s), goal awareness enables the system to focus on achieving target objectives extracted from intents. The proposed approach aligns actions with specific intents, such as improving throughput or reducing energy consumption, making the system pursue operator-defined goals. \gls{HDTGA} can significantly optimize the selection and orchestration of the network-optimizing applications in \gls{RAN}.

This work differs from previous works through the following four key contributions:
\begin{enumerate}
    \item An end-to-end framework based on \gls{GenAI} has been developed to facilitate intent-driven management in wireless networks. The framework incorporates a fine-tuned \gls{LLM} with \gls{QLoRA} providing a memory-efficient platform that works as a guidance for the operator to provide intents. The fine-tuned \gls{LLM} is also adept in processing intents accurately to guide the intent validation and application orchestration phases of the framework.  
    \item A predictive intent validation mechanism has been introduced, utilizing the Informer model. The validation mechanism considers three critical metrics: traffic load, power consumption, and packet loss.
    \item The integration of a \gls{RAG} module has been implemented to enhance real-time decision-making with dynamically retrieved network data.
    \item Unlike previous rule-based and RL approaches, a novel algorithm based on hierarchically organized DTs (\gls{HDTGA}) has been proposed. This algorithm enables the orchestration of network applications by pursuing target metrics extracted from the intents as goals.
\end{enumerate}

Extensive experiments have been conducted to demonstrate the effectiveness of the proposed methodology. First, fine-tuning the \gls{LLM} with a custom dataset significantly outperforms the base \gls{LLM} model, achieving a $6\%$ increase in BERTScore and Meteor metrics. Also, a $9\%$ improvement in semantic similarity is observed. We define and explain BERTScore, Meteor, and semantic similarity in Section V of this paper. The proposed intent validation method effectively predicts traffic load, packet drop rate, and power consumption in future time slots and achieves an average of $88\%$ accuracy. It enables the system to preemptively avoid performance-degrading intents. Finally, we compare the performance of the proposed methodology with \gls{HDTGA} against three baseline schemes. The first two baselines utilize an \gls{HRL} algorithm, one with and the other without the intent validation technique \cite{41}. The last baseline is the vanilla \gls{DT} to perform application selection and orchestration. Our proposed approach surpasses all baseline methods, achieving at least a $19.4\%$ increase in throughput, a $54.9\%$ reduction in network delay, and a $48.5\%$ improvement in energy efficiency.

The remaining parts of the paper are organized as follows: Section \ref{s2} discusses the works related to our research conducted in this paper. The network and the system model with problem formulations are presented in Section \ref{s3}. The proposed methodology with a fine-tuned \gls{LLM}-based query and intent processing module, predictive intent validation, and \gls{HDTGA}-based network application control system are covered in Section \ref{s4}. Performance comparison of the proposed method against the baseline algorithms is presented in Section \ref{s5}. Finally, conclusions are drawn in Section \ref{s6}.

\section{Related Work}
\label{s2}

This section briefly reviews the works related to our research. The proposed methodology has three core components: intent processing, validation, and policy generation for intent execution, as explained before. Therefore, this section is organized to discuss the existing works on each topic separately.

\subsection{Intent Processing}
After the emergence of \gls{LLM}s, intent-driven network management in the area of telecommunications is experiencing a new dimension.
Kristina et al. in their work provide a road map on how intent-driven network automation can be achieved via a three-step methodology: intent processing, validation, and execution \cite{2}. Though this road map provides us with a guideline to work with intent-driven network management, their work focuses only on few-shot learning to process intents. 

Following the methodology in \cite{2}, Habib et al. in their work \cite{8} use few-shot learning and prompt engineering to use a lightweight \gls{LLM} to process intents. 

Intent processing via \gls{LLM} in the literature is so far limited to few-shot learning and prompt engineering as presented in \cite{2} and \cite{8}. The main reason behind this is the high cost and time needed to fine-tune a model based on the specific needs of the relevant use cases.

In this work, instead of using few-shot learning-assisted prompt engineering for intent processing, we use a \gls{QLoRA}-based fine-tuned \gls{LLM}. Unlike past works, this approach provides a memory-efficient solution for intent processing. Furthermore, the operator can query the fine-tuned \gls{LLM} to fetch any necessary network-related information that can aid him in providing an intent.    

\subsection{Intent Validation} 

Intent validation is the process of assessing whether the current network conditions and capacity can accommodate a given intent \cite{45}. This involves verifying whether the available resources, traffic conditions, and operational constraints allow for the successful execution of the intents.

Intent validation is a crucial part of ensuring robust intent-driven automation in the network. Therefore, almost all the works in the literature attempt to validate intents. These works are often limited to rule-based validation methodologies. For example, in \cite{3,5}, authors use a rule-based approach in its intent validation stage where policies are checked for consistency and correct sequence. However, these approaches lack the adaptability to anticipated network changes and need to be updated each time a new intent or service is added.

Considering the limitations of these works, Habib et al. presented a method for intent validation in a predictive manner \cite{8}. However, their methodology is limited to only predicting traffic load, while other crucial parameters can aid in more precise intent validation.

In this work, rather than relying on traditional rule-based validation methods, we propose a predictive intent validation mechanism. Our approach utilizes a transformer-based time-series forecasting model (Informer) to anticipate key network parameters such as traffic load, packet loss, and power consumption. This enables the system to evaluate the feasibility of the intents before execution to ensure that they are aligned with the current network status.

\subsection{Intent Execution via Application Orchestration}

Static optimization-based approaches used in the literature for application orchestration \cite{14} may need recomputation when network conditions change. \gls{RL} is more suitable for dynamic environments compared to optimization-based approaches because it can continuously learn from feedback. Considering this fact, Habib et al. proposed an HRL-based application orchestration methodology in \cite{7,8}. However, these approaches rely highly on hand-crafted reward functions making them hard to generalize.

To overcome the algorithmic limitations of the \gls{RL} methods presented in \cite{7} and \cite{8}, in this paper, we use \gls{DT} to optimize the selection and orchestration of network applications for intent-driven network management. Next, we review the existing works associated with \gls{DT} in wireless in this section.  

To the best of our knowledge, two prior works have used \gls{DT} in wireless scenarios. \gls{DT} has been utilized to provide energy saving in cellular networks by controlling cell on-off switching \cite{15}. The approach presented in \cite{15} improves energy efficiency and throughput by predicting cell activation patterns based on historical traffic data. Another work proposes a \gls{DT} architecture for adaptive wireless resource management \cite{16} where a \gls{DT} is pre-trained on cloud servers with data from various network scenarios and fine-tuned on edge devices.

In this work, we propose a new hierarchical algorithm based on \gls{DT}s named \gls{HDTGA}. Unlike conventional \gls{DT}s used in \cite{15,16}, which rely on predefined returns-to-go, \gls{HDTGA} introduces a goal-aware hierarchical structure that dynamically identifies critical past actions to optimize decision-making. This design enables more efficient intent fulfillment. \gls{HDTGA} can be an excellent fit for network optimization problems since it can use historical context and predictive insights to optimize actions dynamically. 

To showcase the difference between the proposed method and the existing works, we provide Table \ref{comparison}.

\begin{table*}[h!]
\centering
\caption{Feature Comparison Between the Proposed Method and Existing Works}
\begin{tabular}{@{}cccccccc@{}}
\toprule
\textbf{Ref.} & \makecell{\textbf{Intent} \\ \textbf{processing} \\ \textbf{using \gls{LLM}}} & \makecell{\textbf{\gls{LLM}-based network} \\ \textbf{assistance for intent} \\ \textbf{provisioning}} & \makecell{\textbf{Memory-efficient} \\ \textbf{use of \gls{LLM}}} & \makecell{\textbf{\gls{RAG}} \\ \textbf{module}} & \makecell{\textbf{Predictive} \\ \textbf{intent validation} \\ \textbf{using traffic load,} \\ \textbf{packet loss,} \\ \textbf{power consumption}} & \makecell{\textbf{Intent execution} \\ \textbf{via network} \\ \textbf{Applications}} & \makecell{\textbf{Application} \\ \textbf{orchestration} \\ \textbf{using} \\ \textbf{decision} \\ \textbf{transformer}} \\ \midrule
\cite{2}  & Yes & Yes & No  & No  & No  & No  & No  \\
\cite{3}  & Yes & Yes & No  & No  & No  & Yes & No  \\
\cite{5}  & Yes & Yes & No  & No  & No  & Yes & No  \\
\cite{8}  & Yes & No  & No  & No  & No  & Yes & No  \\
\cite{14} & No  & No  & No  & No  & No  & No  & No  \\ 
\textbf{Proposed}  & \textbf{Yes}  & \textbf{Yes}  & \textbf{Yes}  & \textbf{Yes}  & \textbf{Yes} & \textbf{Yes}  & \textbf{Yes}  \\
\bottomrule
\end{tabular}
\label{comparison}
\end{table*}

\section{System Model and Problem Formulation}

\label{s3}

In this section first, we elaborately discuss our system model. Next, we formulate the problems addressed by our three-step methodology, beginning with intent processing, followed by intent validation, and concluding with application orchestration for intent execution. 

\subsection{System Model}
\subsubsection{Network Elements}

An \gls{OFDM}-based cellular system is considered in this work where multiple $B$ serve $U$ users. $B$ represents a \gls{BS} in this paper. Multiple small cells are deployed within the coverage area of a macro cell. The system supports $K$ traffic classes, and users maintain connections with two different \gls{RAT} using dual connectivity. The wireless system model used in this work is illustrated in Fig. \ref{fig1}.

We use \gls{CDL} models (A, B, C, and D) to represent realistic channel propagation characteristics \cite{31}. Each \gls{CDL} model consists of clusters of multipath components, where each cluster represents a physical scatterer. Within each cluster, individual rays have specific delays, angles, and power levels, capturing multipath effects for both \gls{LOS} and \gls{NLOS} conditions. The \gls{CDL} channel response for a user $u$ connected to base station $b$ on subcarrier  $\psi$ can be modeled as the superposition of all multipath components within the clusters:
\begin{equation}
h_{u,b}^{\psi}(t) = \sum_{l_c=1}^{L_c} \sum_{m=1}^{M_{l_c}} \text{amp}_{l_c,m}^{\psi} \, e^{j\theta_{l_c,m}} \, \delta\left(t - \chi_{l_c,m}\right),    
\end{equation}
where $L_c$  is the number of clusters (each cluster represents a physical path), $M_{l_c}$ is the number of rays within the $l_c$-th cluster, $\text{amp}_{l_c,m}^{\psi}$ is the complex amplitude for the $m$-th ray in the $l_c$-th cluster on subcarrier $\psi$, $\theta_{l_c,m}$ is the phase shift for each ray, modeled as a random variable, $\chi_{l_c,m}$ is the delay of the $m$-th ray in the $l_c$-th cluster, $\delta(\cdot)$ is the Dirac delta function representing the arrival time of each ray.

The power delay profile for the channel model is given by the exponential decay of power with delay for each cluster:
\begin{equation}
    P(\chi_{l_c,m}) = P_{l_c} e^{-\chi_{l_c,m} / \sigma_\chi},
\end{equation}
where $P_{l_c}$ is the initial power of the $l_c$-th cluster,
$\sigma_\chi$ is the delay spread for each cluster.

For a user $u$ connected to a \gls{BS} $b$ on subcarrier $\psi$, the channel capacity $\xi_{\psi}$ can be expressed as:
\begin{equation}
\xi_{\psi} = B_w \log_2 \left( 1 + \frac{P_{sc}}{N_0 B} |h_{u,b}^{\psi}(t)|^2 \right),
\end{equation}
where $B_w$ is the bandwidth of each subcarrier, $P_{sc}$ is the transmit power allocated to subcarrier $\psi$, $N_0$ is the noise spectral density, $h_{u,b}^{\psi}(t)$ is the channel response for subcarrier $\psi$ at time $t$.

The proposed system model defines delay as the sum of transmission delay and queuing delay

\subsubsection{Controller Architecture} 
\begin{figure}[!t]
\centerline{\includegraphics[width=0.7\linewidth]{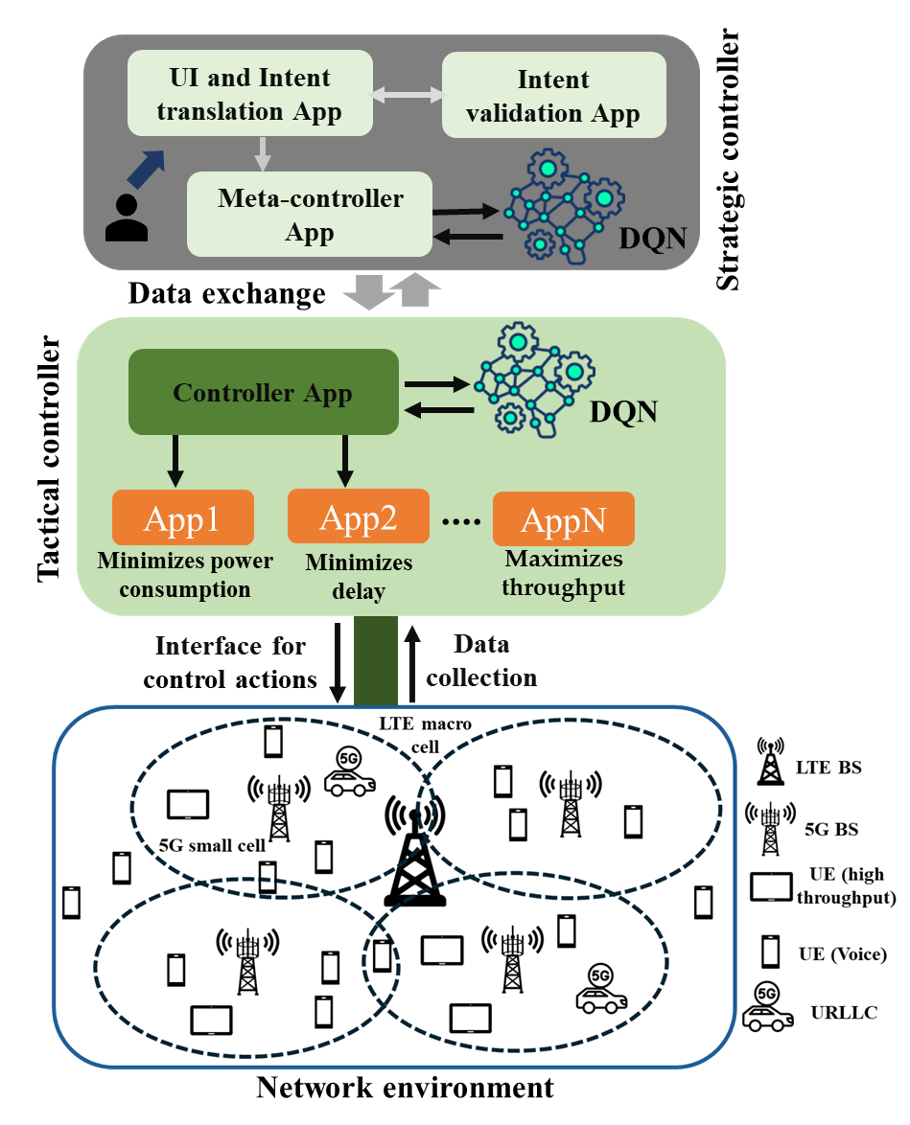}}
\caption{Network and system model.}
\label{fig1}
\vspace{-1.2em}
\end{figure}

The system model, shown in Fig. \ref{fig1}, proposes two distinct controllers responsible for managing RAN functionalities: the strategic controller and the tactical controller. The strategic controller on top is responsible for long-term goals, such as network-wide performance optimizations and policy formulation. It can define the network’s broader objectives, including service quality guarantees, energy efficiency, or overall traffic management for a timespan. The tactical controller on the bottom focuses on real-time decisions, such as handling immediate network changes, performing handovers, or dynamically adjusting resources based on the current network state. It operates within the framework set by the strategic controller but is more focused on near-real-time adjustments to ensure smooth network operation from moment to moment.

\subsubsection{Applications}The controllers host applications focusing on control and optimization tasks across two different time scales. We deploy \gls{BS}s capable of switching bands from 5G mid-band to high-band frequencies \cite{27}. This deployment enables us to cater to high-throughput traffic by employing intelligent beamforming techniques. An application can be developed to regulate power based on \gls{UE} location, utilizing minimal transmission power for enhanced energy efficiency. The system facilitates analog beamforming \cite{11}. The beamforming weights for each beamforming vector are implemented using constant modulus phase shifters. A beam steering-based codebook is considered, from which beamforming vectors are selected \cite{11}. Each BS $b$ transmits with power $P_{TX,b} \in P$, where $P$ denotes the set of possible transmit power levels. The power consumption model for the BS is taken from \cite{12}.

\subsection{Problem formulation}

\subsubsection{Intent and Query Processing by \gls{LLM}}
The first step of the proposed methodology is to fine-tune an \gls{LLM} using custom datasets. We fine-tune an \gls{LLM} so that it can (i) respond to human queries about the system (e.g., reporting current network load, throughput, latency, and so on.) and (ii) process explicit performance-oriented intents (e.g., ``Increase throughput by 10\%''). These two tasks are referred to as:
\begin{enumerate}
    \item \textit{subproblem 1}: Query-Response
    \item \textit{subproblem 2}: Intent Processing
\end{enumerate}

In \textit{subproblem 1}, the \gls{LLM} is trained to generate correct responses to domain-specific queries. In \textit{subproblem 2}, the \gls{LLM} is trained to extract two key components from an intent: (a)~the performance metric (e.g., ``throughput,'' ``latency,'' ``energy efficiency''), and (b)~the requested magnitude of change (e.g., ``10\%'', ``5\%''). Utilizing the capability of an \gls{LLM} to handle complex natural language, it is possible to jointly address both tasks by fine-tuning on labeled data.

Let $\mathcal{D}_{qr} = \{(q_j, r_j)\}_{j=1}^{F}$ be a dataset of $F$ query--response pairs where $q_j$ is the $j$-th query (e.g., ``What is the current throughput?'') and $r_j$ is the ground truth response to the query $q_j$.

We fine-tune the \gls{LLM} (denoted by $f_{\theta}$, with parameters $\theta$) on $\mathcal{D}_{qr}$ so that for any query $q_j$, the model's predicted response $\hat{r}_j = f_{\theta}(q_j)$ aligns well with the corresponding ground truth $r_j$. A standard approach is to minimize a loss function $\mathcal{L}(r_j, \hat{r}_j)$ that measures the difference between the predicted and true responses. The \gls{LLM} learns how to generate the correct response for each query by adjusting its parameters via supervised training.

For the \textit{subproblem 2}, let us assume $I = \{i_1, i_2, \ldots, i_n\}$ to be the set of intents and $I_{t_y} = \{t_{y_1}, t_{y_2}, \ldots, t_{y_e}\}$ be the set of intent types. We define a function $f: I \rightarrow I_{t_y}$ that maps each intent to an intent type. The system allows multiple intents to map to the same type as well as single intents mapping to one type. This function $f(i)=t_y$ indicates that the intent $i$ belongs to the intent type $t_y$. Each intent $i \in I$ is mapped to a type $t_y \in I_{t_y}$. 

Furthermore, we also define a relation $R \subseteq I \times I_{t_y}$ that relates intents to their types, accommodating both one-to-one and many-to-one mappings. It is represented as follows: 
\begin{equation}
    R=\{(i,t_y)|i\in I, t_y \in I_{t_y} \text{ and }i \text{ is of type } t_y \} 
\end{equation}

This allows for pairs $(i,t_y)$ where different $i \in I$ can correspond to the same $t_y \in I_{t_y}$, and also allows for single intents mapping to single types. 

Let $\{I_{s_1}, I_{s_2}, \ldots, I_j\}$ be the subsets of $I$ that are disjoint. This means the intersection of any two subsets results in an empty set. For example: $I_{s_3} \cap I_{s_5} = \varnothing$. Each subset of intents can be classified into distinct intent types from the set $I_{t_y}$. 

\begin{equation}
    I = \bigcup_{j=1}^{N} I_{s_j}.   
\end{equation}

Each intent $i_j \in I$ is parsed to identify:
\begin{itemize}
    \item Type($t_{y_j}$): example: ``throughput,'' ``latency,'' ``energy efficiency.''
    \item Magnitude($\Lambda_j$): example: ``10\%'' or ``5\%.''
\end{itemize}

We define $\mathcal{D}_{intents} = \{(i_j, t_{y_j}, \Lambda_j)\}_{j=1}^N$ as the dataset of $N$ labeled intents. Here, $i_j$ is the $j$-th intent text, $t_{y_j}$ is the true ``type'' label, and $\Lambda_j$ is the true ``magnitude'' label. After fine-tuning, the LLM will predict $\hat{t}_{y_j} = f_{\theta}^{\text{type}}(i_j)$ and $\hat{\Lambda}_j = f_{\theta}^{\text{mag}}(i_j)$.

During supervised training, these predictions are compared to the ground truth $(t_{y_j}, \Lambda_j)$. A typical training setup involves loss functions $\mathcal{L}_{\text{type}}$ and $\mathcal{L}_{\text{mag}}$, which measure how accurately the model identifies the correct type and magnitude. The model parameters $\theta$ are updated so as to improve classification accuracy for both type and magnitude.

To address both subproblems in a single framework, we jointly fine-tune the \gls{LLM} on $\mathcal{D}_{qr}$ and $\mathcal{D}_{intents}$. By exploiting the \gls{LLM}'s robust \gls{NLP} capabilities, the model can simultaneously learn to respond accurately to system-related queries and identify intent types (e.g., throughput, latency, energy efficiency) and their corresponding magnitude requests. 

\subsubsection{Intent Validation}
The second problem of the proposed framework is associated with intent validation. Let $V = \{v_{c_1}, v_{c_2}, \ldots, v_{c_n}\}$ be the set of validation checks, and $O = \{o_1, o_2, \ldots, o_n\}$ be the sequence of policies. Our objective function in this case can be formulated as follows: 
\begin{equation}
    \min \sum_{i=1}^n \sum_{o=1}^n \vartheta_{i,o},
\end{equation}
where $\vartheta_{i,o}$ represents the performance degradation score because of policies passing validation checks. Each policy must go through a validation check $v_{c_i}$. We calculate this score based on \gls{QoS} drifts. QoS parameters deviate from the originally defined \gls{QoS} metrics over time due to performance degradation in the wireless networks causing \gls{QoS} drifts. By deviation, we mean when a \gls{QoS} parameter has a lesser value than the required. For a specific traffic class, $t_{class}$ having a set of \gls{QoS} requirements $Q_{req}$ and associated performance metrics $P_{metric}$, there can be \gls{QoS} drifts for multiple different \gls{QoS} definitions.
\begin{equation}
Q_{\text{drift}}(t) \;=\;
\begin{cases}
\max\!\bigl(0,\;A_{P_{metric}}(t)\;-\;D_{\text{QoS}}\bigr), 
& \text{for $P_{metric(l)}$},\\[8pt]
\max\!\bigl(0,\;D_{\text{QoS}}\;-\;A_{P_{metric}}(t)\bigr),
& \text{for $P_{metric(h)}$}.
\end{cases}
\label{qdr}
\end{equation}

Here, $A_{P_{metric}}$ represents the achieved value of a metric for a certain traffic type at time $t$, and $D_{\text{QoS}}$ represents the pre-defined value of the same parameter which has to be maintained for optimal performance. $P_{metric(l)}$ refers to a performance metric for which lower values are better for the system and QoS maintenance. On the other hand, $P_{metric(h)}$ refers to a performance metric for which higher values are preferred. 

Intents that do not cause any \gls{QoS} drift are validated. Intents are not allowed to change the network configuration if they will lead to \gls{QoS} drifts to avoid drastic performance degradation in the system.  

\subsubsection{Application Orchestration}

Let $A$ be the set of all the applications in our \gls{IRCs}. By \gls{IRCs} we denote the strategic and tactical controllers presented in Fig. \ref{fig1}. For each IRC application $a\in A$, $\mathcal{F}_a \subseteq \mathcal{F}$ represents the subset of functionalities offered by $a$. Accordingly, we define a binary variable $\varrho_{a,f}\in\{0,1\}$. This variable is $1$ if $f\in\mathcal{F}_a$. It is $0$ otherwise. Two applications (can either be an application running on a strategic or tactical controller) $a_1$ and $a_2$ can provide similar functionalities, i.e., $\mathcal{F}_{a_1} = \mathcal{F}_{a_2}$. However, they can differ in terms of model, architecture as well as required inputs. For example, two applications $a_1$ and $a_2$ may provide the same functionality of increasing energy efficiency but one application may use \gls{DRL} and another may use a supervised approach.

The resource requirement for deploying and running application $a$ is denoted by $r^\omega_a$, where $\omega \in \Omega$ represents the resource type. Let $\mathcal{T}$ be the set of all possible input types. Each application $a \in A$ requires a specific input type, represented as $t^{IN}_a \in \mathcal{T}$ (e.g., Traffic flow type, \gls{SINR}, and queue length measurements). Furthermore, for every intent $i$, we specify the set of functionalities that an application must perform at a designated location $l$. 

Certain functionalities impose strict latency constraints, making it inefficient or impractical to execute them on nodes located far from where the input is generated. To address this, we define $\delta_{i,f,l} \geq 0$ as the maximum latency an intent $i$ can tolerate when executing application $a$ at location $l$. Given resource limitations, an application $a$ can be deployed up to $C_{a,l}$ times at a specific location ($l$), such as Controllers, CU, or DU. Additionally, an application can be instantiated multiple times on the same node or location to accommodate different traffic classes. 

Let $\eta^{i,f,l}_{a,\kappa}\in\{0,1\}$ be a binary variable such that $\eta^{i,f,l}_{a,\kappa}=1$ if functionality $f$ demanded by intent $i$ is provided by instance $\kappa$ of application $a$ instantiated on location $l$.

Each application $a$ requires a specific type of input $t^{IN}_a$ and as a consequence, when instantiating an application on any location $l$, we must ensure that such input is available at that very location within the desired time-scale. It is essential to ensure that the time required to gather input from nodes in $\mathcal{D}^{IN}_{i,f,l}$ remains within the allowable latency limit $\delta_{i,f,l}$ for each intent.
For each orchestration policy $\mathbf{Z}$, the \textit{data collection time} can be formalized as follows: 
\begin{align} 
\Delta_{i,f,l}(\mathbf{Z}) = 
\sum_{a \in \mathcal{A}} \varrho_{a,f} \sum_{\kappa=1}^{C_{a,l}} \eta^{i,f,l}_{a,\kappa} \sum_{l,l' \in \mathcal{L}^{\text{IN}}_{i,f}} \mu_{i, f, a,l,l'},
\label{pol1}
\end{align}
where $\mu_{i,f,a,l,l'} = \left(\frac{s_{t^{IN}_a}}{b_{l,l'} |\mathcal{L}^{IN}_{i,f}|}\! + T_{l,l'} \right)$, $s_{t^{IN}_a}$ is the input size of an application $a$ measured in bytes, $b_{l,l'}$ is the data rate of the link between nodes $l$ and $l'$, and $T_{l,l'}$ represents the propagation delay between nodes $l'$ and $l$.   

The latency constrain can be formulated as follows: 
\begin{equation} \label{Latency}
  \Delta_{i,f,l}(\mathbf{Z}) + \Delta^{EXEC}_{i,f,l}(\mathbf{Z}) \leq \delta_{i,f,l}.
\end{equation}
Here, $\Delta^{EXEC}_{i,f,l}$ is the time to execute an application $a$ on a location $l$. We declare a variable combining these two types of latency:
\begin{equation}
  \Delta_{LAT}=\Delta_{i,f,l}(\mathbf{Z}) + \Delta^{EXEC}_{i,f,l}(\mathbf{Z}).
\end{equation}

The problem of application orchestration that we want to address is to overcome the excessive latency that can be an issue while executing an intent. To deal with it we consider the following ratio: 
\begin{equation}
  \Delta{(o_i)}= \frac{\Delta_{LAT}}{\delta_{i,f,l}}. 
\end{equation}

Similarly, we take the ratio of the total used memory and the available memory ($ r^ {\omega_1}_{a}= \frac{m_u}{m_{max}}$), the ratio of the total used processing power and the available processing power ($r^ {\omega_2}_{a}= \frac{p_u}{p_{max}}$) for executing an intent. The computing resource consumption is the summation of these two types of resources for executing a policy resulting due to an intent.  
\begin{equation}
  R ({o_i}) = r^ {\omega_1}_{a} + r^ {\omega_2}_{a}. 
\end{equation}

We define a utility function as follows:  
\begin{equation}  
    U(o_i) = |M_p| - \varpi \gamma_{q_s},  
\end{equation}  
where $M_p$ denotes the magnitude of the performance metric that the operator aims to enhance. The term $\varpi$ serves as a penalty parameter for violations of \gls{QoS} requirements, while $\gamma_{q_s}$ represents the number of \gls{UE}s experiencing \gls{QoS} violations. These violations are evaluated based on the \gls{QoS} parameters associated with each \gls{UE}’s traffic class.  

Finally, the third problem we want to address in our proposed three-step methodology is represented by
\begin{align}
    \underset{\mathbf{Z}}{\max}\ &\ \sum_{i=1}^n \sum_{t_y=1}^e \bigg[ U(o_i) - g_1 \times R(o_i) - g_2 \times \Delta(o_i) \bigg], \label{eq:25} \\
    \text{s.t.} \quad & (\ref{Latency}), \tag{14a}\label{eq:25a} \\
    & \sum_{a \in \mathcal{A}} \sum_{\kappa=1}^{C_{a,l}} \eta^{i,f,l}_{a,k} \cdot m_u \leq m_{\text{max}}, \quad \forall l, \tag{14b}\label{eq:25b} \\
    & \sum_{a \in \mathcal{A}} \sum_{\kappa=1}^{C_{a,l}} \eta^{i,f,l}_{a,k} \cdot p_u \leq p_{\text{max}}, \quad \forall l, \tag{14c}\label{eq:25c} \\
    & \Delta_{i,f,l}(\mathbf{Z}) \leq \delta_{i,f,l}, \quad \forall i, f, l, \tag{14d}\label{eq:25e} \\
    & \gamma_{q_s} \leq Q_{\text{max}}, \tag{14e}\label{eq:25f} \\
    & \varrho_{a,f} \in \{0, 1\}, \quad \eta^{i,f,l}_{a,\kappa} \in \{0, 1\}, \quad \forall a, f, l, \kappa, \tag{14f}\label{eq:25g}
\end{align}
where, $g_1$ and $g_2$ are the weight factors. Among the constraints, (\ref{eq:25b}) ensures that the memory resources required ($m_u$) by all instances of applications on a location \( l \) do not exceed the available memory capacity \( m_{\text{max}} \) at that location. (\ref{eq:25c}) makes sure that the processing power used by all instances of applications on a location \( l \) does not exceed the available processing power \( p_{\text{max}} \). (\ref{eq:25e}) ensures that each application \( a \) requiring a specific input type \( t^{IN}_a \) has that input available at the location \( l \) within the desired time-scale. (\ref{eq:25f}) ensures that the number of UEs experiencing QoS violations does not exceed the predefined threshold \( Q_{\text{max}} \). Lastly, (\ref{eq:25g}) enforces that binary variables \( \varrho_{a,f} \) and \( \eta^{i,f,l}_{a,\kappa} \) take values only in \(\{0, 1\}\), representing the selection of functionalities and instances. Problem formulated in (\ref{eq:25}) is the last problem formulation of our methodology which we would refer to as Problem 3 throughout the paper. (\ref{pol1}) and (\ref{Latency}) are adopted from the idea and formulations from \cite{14}. 

The problem in (\ref{eq:25}) presents complexities that challenge classical optimization methods. The utility function $U(o_i)$ includes penalties for QoS violations and intended performance metrics, which are often non-linear and non-convex. Classical optimization struggles with such objective functions, especially when the feasible region is non-convex. Additionally, the problem requires dynamic inputs (e.g., traffic flow, \gls{SINR}, queue length) that vary with orchestration policy $\mathbf{Z}$, making it inherently stochastic and context-sensitive. These factors render classical techniques impractical, especially for large-scale systems with numerous applications and UEs. To address this, we employ RL, specifically \gls{DT}s, which offer approximate solutions efficiently. DTs are well-suited for real-time or near-real-time decision-making in IRCs, dynamically adapting to evolving network conditions.   

\section{Proposed Three-Step Methodology for Intent-driven Network Management}
\label{s4}

\begin{figure*}[!t]
\centerline{\includegraphics[width=0.9\linewidth]{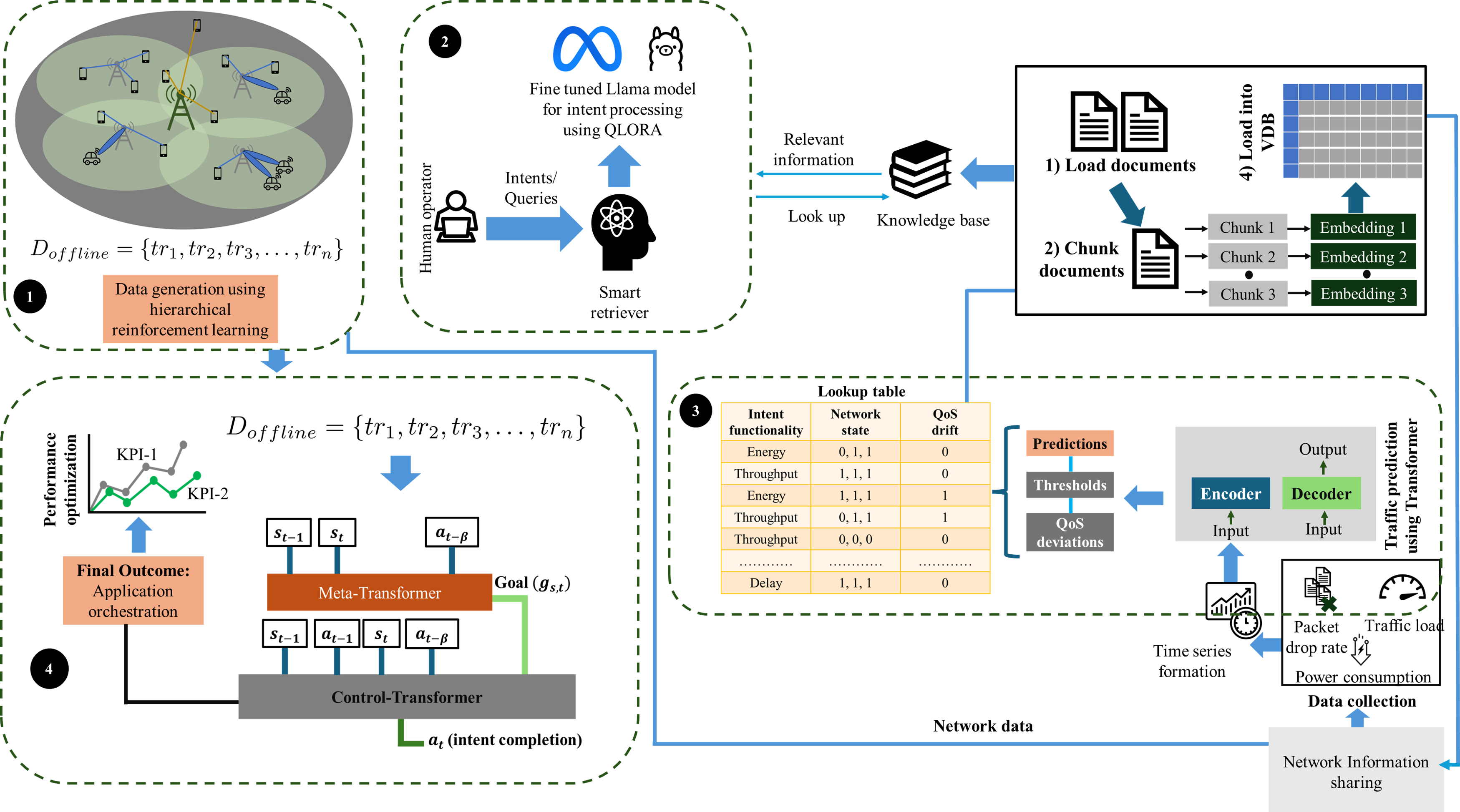}}
\caption{Step-by-step illustration of the proposed methodology.}
\label{SMB}
\vspace{-1.2em}
\end{figure*}

In this section, we extensively discuss the three-step methodology of intent and query processing, intent validation, and application orchestration in RAN. However, these three steps at first require data from a network infrastructure. Based on the network and the system model presented in the previous section, we run extensive simulations. Simulations included providing the system with intents and running an \gls{HRL} algorithm to generate training data for the proposed \gls{HDTGA} which is based on the \gls{DT}. Fig. \ref{SMB} presents the proposed methodology at a glance. We use numerals in black circles to portray the steps of the proposed method. Generating training data is marked as the first step. Intent processing, validation, and application selection with orchestration via \gls{DT} are represented as the second, third, and fourth stages in the figure.  

A \gls{h-DQN} having a two-level hierarchy with a meta-controller on top \cite{18} and a controller on the bottom has been used to generate our training data. These two controllers are solely related to the \gls{h-DQN} algorithm that we use to manage network-optimizing applications. Meta-controller and the controller are hosted in the strategic and tactical controller, respectively. Meta-controller in \gls{h-DQN} can take in a network state (e.g., traffic class) and a goal (desired change of a performance metric extracted from an intent) to achieve. The controller in the lower level takes the action of choosing an application or combinations of them based on the state and the goal. The meta-controller is designed as a top-level application in the strategic controller (see Fig. \ref{fig2}). Applications are managed by the controller which can directly optimize system performance by initiating such applications. 

Let $\mathcal{S}$ be the state space, $\mathcal{A}$ the action space, and $\mathcal{G}$ the goal space. We define the following \gls{MDP} for the \gls{h-DQN} for application selection and orchestration.    

\begin{enumerate}
    \item \textbf{States:} The state space consists of four elements: $\mathcal{S}=\{P_C, t_{class}, T_L, P_L\}$. The first element of the state space is $P_C$, representing power consumption. Including power consumption as a state helps directly manage energy efficiency. The second element is traffic load, represented by $T_L$ in the state space. Including traffic load as a state helps in choosing actions suitable for the overall traffic scenario. The third element is packet loss percentage ($P_L$). It is included in the state space because it is a critical \gls{QoS} metric that directly impacts user experience. By including it in the \gls{MDP}, the system can prioritize actions that minimize packet loss. Lastly, traffic flow types ($t_{class}$) of different users in the network are used as states too. Both meta-controller and controller share the same states.
    \item \textbf{Actions:} Selecting network optimizing applications, or combinations of them is considered as actions to be performed by the controller which is defined as $\mathcal{A}_{\text{h-DQN}}=\{A_{App1}, A_{App1,2},...., A_{AppN}\}$.
    \item \textbf{Intrinsic reward:} The intrinsic reward function ($r_{in}$) for the controller is: $r_{in}=C_\rho-\varpi \gamma_{q_s}$. Here, $C$ is the magnitude of the intended performance metric the operator intends to improve.
    \item \textbf{Goal for the controller:} Increased level of a performance metric that can satisfy operator intent is passed to the controller as goals. For example, $\mathcal{G}=\{TP_1, TP_2,..., TP_n\}$ for throughput increasing intents.
    \item \textbf{Extrinsic reward:} Summation of the intrinsic reward over $\tau$ steps.
\end{enumerate}

We generate \gls{RL} trajectories through a sequence of steps. First, the operator provides an intent to improve a specific performance metric, such as increasing throughput by $x\%$ or reducing power consumption by $y\%$, leading to a new target value. Next, the meta-controller application assigns these targets as goals for the controller. The controller then selects one or more applications to achieve the target as closely as possible. The system learns by receiving rewards based on how effective the selected applications are. Finally, the chosen applications, using \gls{DRL}-based functionalities, optimize network performance in response to the operator's intent.

We run network simulations based on the system model and \gls{h-DQN}-based network application selection described previously and record trajectories as an offline dataset. We represent this dataset as follows: 
\begin{equation}
    D_{offline} = \{tr_1,tr_2,tr_3,\ldots ,tr_n\},
\end{equation}
where $tr_n=\{g_n,s_n,a_{t_n}\}$. We denote goals, states, and actions as $g$, $s$, and $a_t$ respectively. These data will be used at the later stage of the work to train the \gls{HDTGA} algorithm for near-real-time application orchestration. Running the network infrastructure will also provide us with some crucial information about the system. For example, we can record what is the incoming traffic load or what are the available applications we have in the inventory for network optimization. This whole process in Fig. \ref{SMB} is marked as ``1" in a black circle.  

\subsection{\gls{QLoRA}-based fine-tuned \gls{LLM} for Query and Intent Processing}

In this subsection of the paper, we fine-tune an \gls{LLM} model using a custom dataset of query-response pairs and intents. A human operator can query for any kind of relevant network information that can be helpful in providing an intent.

A dataset is built using the data we collected from the simulations. The dataset can be divided into three segments, each presenting a distinct type of data or interaction within the system.
\begin{enumerate}
    \item \textbf{Dynamic Network Information:} This segment presents examples of real-time data extracted from the network. Such information changes dynamically over time as the network evolves.
    \item \textbf{Queries and Responses Associated with Semi-Static Information:} The second segment highlights prerecorded, semi-static information that can be accessed via specific queries. This type of information remains constant if it is not explicitly modified by the system administrator.
    \item \textbf{Intent Prompts and Responses:} The final segment provides examples of the intent prompts along with the corresponding responses. This segment is crucial for the later stages of the methodology (policy generation).  
\end{enumerate}

\begin{figure}[!t]
\centerline{\includegraphics[width=1\linewidth]{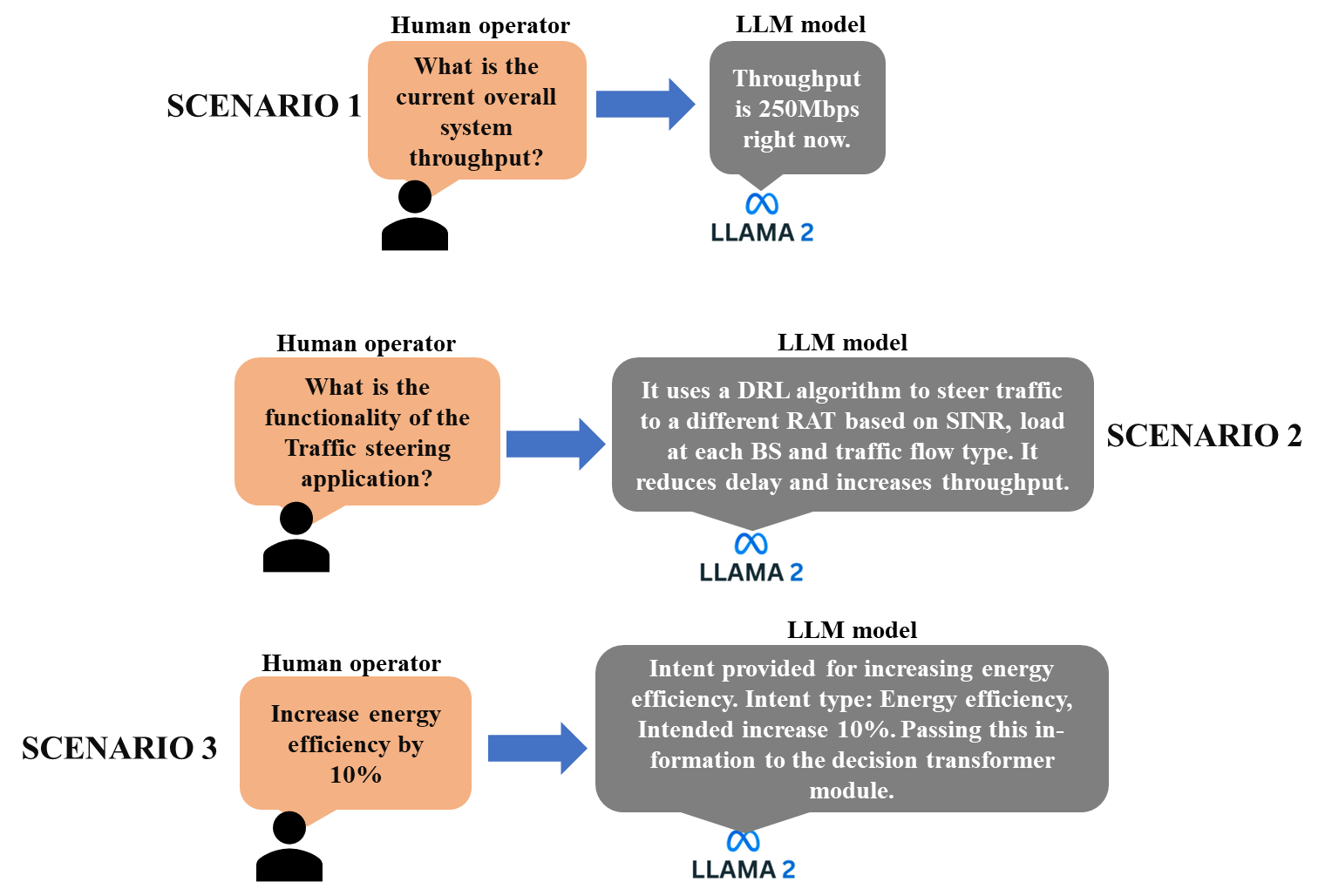}}
\caption{Example queries/intents with responses.}
\label{fig2}
\vspace{-1.2em}
\end{figure}

Fig. \ref{fig2} represents three scenarios of query and intent processing via an \gls{LLM}. Using similar patterns of information, queries, and responses, hundreds or even thousands of query-response pairs can be generated. This is the kind of dataset used for fine-tuning the \gls{LLM} in this research.  

Based on the custom dataset, we train the \gls{LLM} using \gls{QLoRA}. The problem QLoRA addresses is the need to fine-tune LLMs while using limited computational resources. Fine-tuning large models in 16-bit precision requires significant memory and high-end hardware. \gls{QLoRA} offers a solution by reducing memory usage and enabling the fine-tuning of large models on a single \gls{GPU}.

\gls{QLoRA} introduces four key techniques for efficient model fine-tuning. The first is one is \textit{4-bit Normal Float Quantization}. It reduces memory usage by quantizing model parameters into 16 discrete ``buckets''. Unlike standard quantization, which poorly represents values near zero, this approach uses equally sized buckets to optimize parameter distribution. It balances memory efficiency and fidelity by ensuring better approximation of high-density regions.  

The second technique is \textit{Double Quantization}. It addresses the limitations of standard quantization by employing a blockwise quantization approach. In conventional quantization, parameter values are mapped into a fixed range, such as $[-127, 127]$, using the rescaling formula:  

\begin{equation}
    P_{q}(\varsigma_i) = \left\lfloor \frac{\varsigma_i}{\text{max}(|\mathbf{X}|)} \times 127 \right\rfloor.    
\end{equation}

However, extreme values in the tensor $\mathbf{X}$ can skew this process. To mitigate this, blockwise quantization splits $\mathbf{X}$ into $\mu$ smaller blocks $\{X_1, X_2, \ldots, X_\mu\}$, each independently scaled as:  

\begin{equation}
    P_{q}(X_\mu) = \left\lfloor \frac{X_\mu}{\text{max}(|X_\mu|)} \times 127 \right\rfloor.
\end{equation}
This approach prevents outliers from distorting the entire quantization process.  

The third technique associated with \gls{QLoRA} is \textit{Paged Optimization}. It allows offloading parts of the model to the \gls{CPU} and retrieving them when needed if the \gls{GPU} memory capacity has exceeded. This results in efficient training on resource-limited hardware.  

Lastly, \textit{\gls{LoRA}} \cite{32} is one of the most integral parts of \gls{QLoRA} which fine-tunes a pre-trained model by introducing a small, trainable set of parameters. Rather than updating all model parameters, \gls{LoRA} keeps the primary model weights $\mathbf{W}_0$ frozen and learns a low-rank delta matrix, $\Delta \mathbf{W}$ as the product of two smaller matrices $\mathbf{A}$ and $\mathbf{B}$:
\begin{equation}
\Delta \mathbf{W} = \mathbf{A} \cdot \mathbf{B},
\end{equation}
where $\mathbf{A} \in \mathbb{R}^{d \times r}$ and $\mathbf{B} \in \mathbb{R}^{r \times \upsilon}$ with $r \ll d, \upsilon$. The notations $\mathbb{R}^{d \times r}$ and  $\mathbb{R}^{r \times \upsilon}$ indicate the dimensions of the matrices $\mathbf{A}$ and $\mathbf{B}$ in \gls{LoRA}. The fine-tuned model’s weights become:
\begin{equation}
\mathbf{W} = \mathbf{W}_0 + \Delta \mathbf{W} = \mathbf{W}_0 + \mathbf{A} \cdot \mathbf{B}.
\end{equation}
This approach effectively reduces the number of trainable parameters, allowing for efficient memory usage during fine-tuning. 

Together, these four techniques make \gls{QLoRA} a powerful and efficient framework for fine-tuning large language models with minimal computational overhead.

When fine-tuning an \gls{LLM} base model, we utilized all four components of \gls{QLoRA} (quantization, double quantization, paged optimization, and \gls{LoRA}) to optimize for memory and computational efficiency. The training process involved the following steps:
\begin{enumerate}
    \item Performing a forward pass on the model (using the 4-bit quantized weights).
    \item Applying double quantization to mitigate the memory overhead of quantization constants.
    \item Using paged optimization to manage memory between CPU and GPU efficiently.
    \item Updating only the parameters in \(\mathbf{A}\) and \(\mathbf{B}\), as specified by LoRA, reducing the optimizer state’s memory footprint.
\end{enumerate}

For instance, a 10-billion parameter model that would typically require over 160 GB of memory can be fine-tuned using QLoRA with approximately 40 GB. This  process enables scalability even on limited hardware.

\begin{figure}[!t]
\centerline{\includegraphics[width=0.7\linewidth]{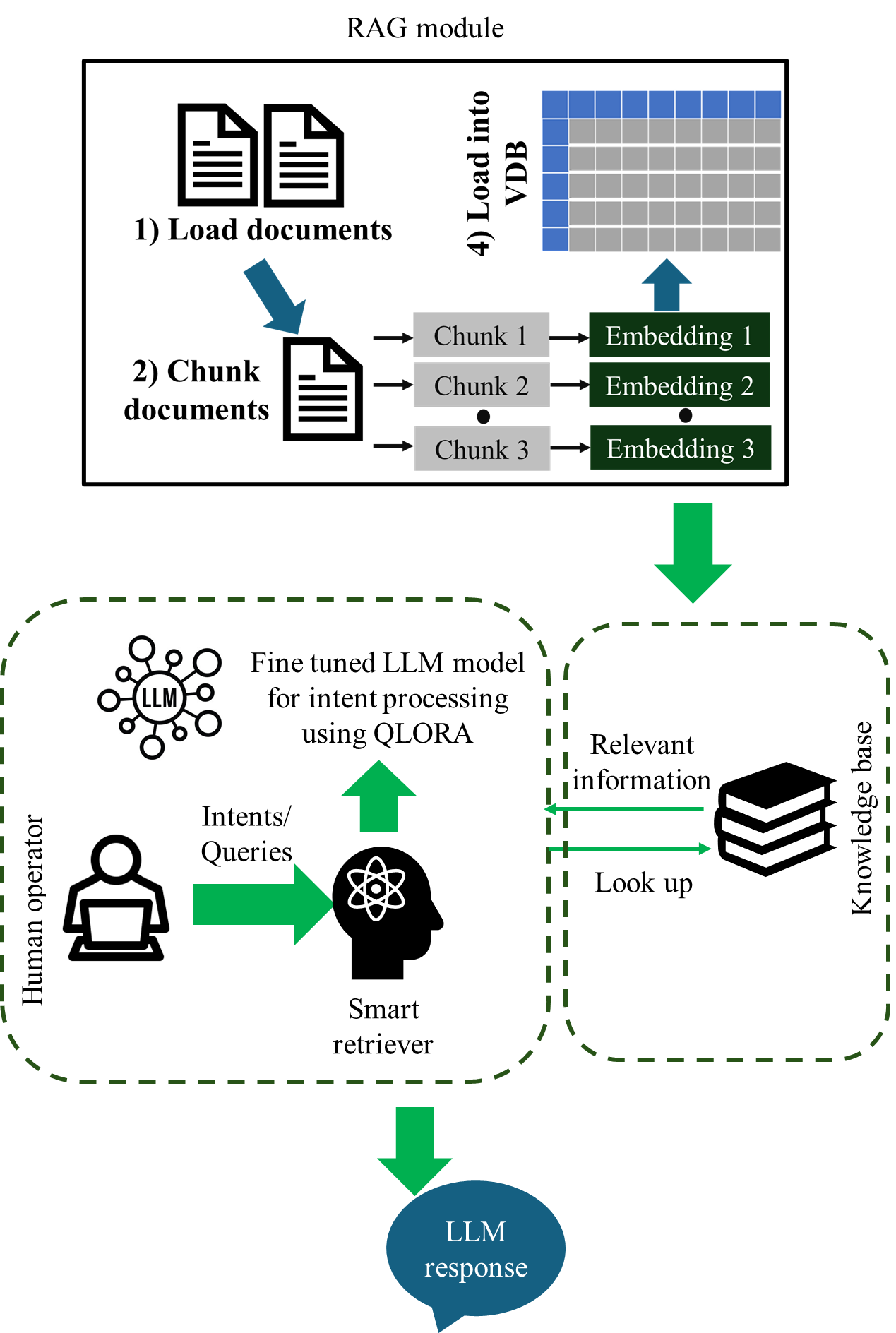}}
\caption{Fine-tuned LLM model with RAG.}
\label{fig3}
\end{figure}

Dataset curated for fine-tuning the \gls{LLM} in this work can be hard to handle via an \gls{LLM} because of its temporal nature as it is supposed to change over time. As we move forward in time, the data will be outdated. To tackle such an issue, we use a \gls{RAG} module \cite{19}. It is a hybrid model architecture that combines retrieval-based methods with generative models to answer queries by accessing external knowledge bases or documents during the response generation process. \gls{RAG} is especially useful when the generative model's internal knowledge is limited, outdated, or incomplete. Given a query, a retriever model fetches relevant documents or passages from an external database, such as a corpus of texts or a custom knowledge base. The retrieved documents are then passed to the generative model along with the query. The generative model processes the query and the relevant information from the retrieved documents.  It uses both the original query and the retrieved context to generate a response, typically combining knowledge from the retrieved documents to provide an accurate and coherent answer. Fig. \ref{fig3} summarizes the whole process of fine-tuning an LLM model (LlaMa \cite{21} is provided as an example in the figure) aided by a RAG module. 

\subsection{Predictive Intent Validation}

At this stage of the work, we present how to validate intents from a human operator. Let us explain with an example. For instance, when there is a high traffic load, an intent to increase energy efficiency can cause performance degradation. To avoid such effects, we perform intent validation using predictions of the future network states. Predictive intent validation helps ensure that network optimization actions (e.g., initiating or terminating applications) are aligned with the expected network conditions. This alignment is crucial for maintaining and enhancing performance metrics like throughput and energy efficiency. In this work, we perform a transformer-based prediction of three crucial parameters: upcoming cell aggregate traffic (traffic volume), packet loss percentage, and power consumption. To perform such predictions, we collect the data using the network's \gls{PDCCH}. Readers are referred to \cite{17} for more details regarding the data collection procedure. Using the data collection methodology in \cite{17}, the time series for predicting $P_L(\%)$ can be formulated as follows: 
\begin{equation}
    P_L(T)=\{P_L(t_1),P_L(t_2),P_L(t_3),\ldots,P_L(t_n)\}.
\end{equation}

Similarly, time-series data for power consumption ($ P_{in} (T)$) and aggregate cell traffic load ($T_{L} (T)$) are also collected. For one-step prediction, a many-to-one architecture \cite{17} is employed. $P_{in}(T)$, $T_{L} (T)$, and $P_L(T)$ are monitored over a predetermined number of time slots up to $T$ and then they are predicted at the subsequent time slot, $T+1$.

We use a novel transformer architecture \cite{11} named Informer for making predictions.  Informer is designed to handle lengthy time-series data. Traditional methods struggle with increasing sequence length and perform poorly. To the best of our knowledge, this is the first work to utilize this architecture for network prediction. 

The input to the model is represented as a time-series sequence. Each time step includes both local positional context (e.g., position embeddings) and global time stamps (e.g., \gls{TTI} in our case) to enhance sequence alignment. The transformer’s self-attention mechanism compares queries, keys, and values using dot-products. Informer optimizes this by computing attention only for the ``dominant" queries based on their sparsity. This reduces the complexity from $O(L^2)$ to $O(L\: log \: L)$ \cite{11}. Here, $L$ represents the sequence length.  

Informer further improves efficiency by progressively condensing attention maps, halving sequence length at each encoder layer in a pyramid-like structure. Unlike autoregressive models, its decoder predicts the full sequence in a single pass using known sequences as start tokens to ensure long-term forecasting without accumulating errors. To further elaborate the prediction process, we present Algorithm 1.

\begin{scriptsize} % Change font size here
\begin{algorithm}
\label{algo1}
\caption{Predicting Traffic Load, Packet Loss Percentage, and Power Consumption using Informer}
\begin{algorithmic}[1] % [1] for showing line numbers
    \State \textbf{Input:} 
    Time series data $\mathbf{D} = \{\mathbf{D_{t_1}}, \mathbf{D_{t_2}}, \ldots, \mathbf{D_{t_n}}\}$
    
    \State \textbf{Data aggregation:} 
    Cell traffic or packet loss (\%) or power consumption 
    \State \textbf{Encoder input:} 
    Past $t_{\text{steps}}$ time steps $\mathbf{X}_{\text{en}}$
    
    \State Process: 
    Apply ProbSparse self-attention mechanism to achieve efficient dependency alignment:
    \[
        A(\text{Query}, \text{Key}, \text{Value)} = \text{Softmax}\left( \frac{\text{Query} \cdot \text{Key}^{Trans}}{\sqrt{d_{\text{step}}}} \right) V
    \]
    \State \textbf{Output:} 
    Encoded representations $\mathbf{H}_{\text{en}}$
    \State \textbf{Decoder input:} 
    Encoded representations $\mathbf{H}_{\text{en}}$ and previous predictions
    \State Process: 
    Apply generative style decoding for long-sequence outputs:
    \[
        \mathbf{Y}_{\text{de}} = \text{Decoder}(\mathbf{H}_{\text{en}}, \mathbf{X}_{\text{prev}})
    \]
    \State \textbf{Output:} 
    Predicted traffic load $\mathbf{P_{L_a} (t+1)}$, packet loss percentage $\mathbf{P_L(t+1)}$, and power consumption $\mathbf{P_{P_c} (t+1)}$
\end{algorithmic}
\end{algorithm}
\end{scriptsize} % End of font size change

As shown in Algorithm 1, the encoder processes a sequence of past time steps of length $t_{\text{steps}}$. It applies the ProbSparse self-attention mechanism \cite{11} to reduce computational complexity by focusing on dominant queries and aligning long-term dependencies efficiently. Self-attention computes dependencies using dot products of queries and keys, weighted by similarity. The encoder generates encoded representations, which the decoder uses along with prior predictions to forecast future values. Unlike autoregressive methods, the decoder predicts entire sequences in a single forward pass which reduces cumulative errors. In Algorithm 1, $\mathbf{Y_{\text{de}}}$ represents the predicted output, while $\mathbf{X_{\text{prev}}}$ is the previous input. This approach enables accurate long-term forecasting of traffic load $\mathbf{P_{L_a} (t+1)}$, packet loss percentage $\mathbf{P_L(t+1)}$, and power consumption $\mathbf{P_{P_c} (t+1)}$ 

Next, we fix some thresholds for these predicted values. For example, $H_{load}$ is the high threshold for $\mathbf{P_{L_a} (t+1)}$, and $L_{load}$ is the low threshold for the same entity. We can define these kinds of thresholds for the predictions associated with power consumption and packet loss percentage too. This leads to a set of thresholds: $S_{th} = \{H_{load}, L_{load}, H_{loss},  L_{loss}, H_{pc},  L_{pc}\}$.

Here, $H_{loss}$, and $L_{loss}$ are the high and low thresholds for packet loss percentage. Furthermore, $H_{pc}$,  $L_{pc}$ are the high and low thresholds for power consumption. Predictions can either fall into the category of a high or a low threshold. If we have an intent: ``Increase throughput by 12\%", the ideal situation of executing such intent would be when the predicted traffic load ($\mathbf{P_{L_a} (t+1)}$) is very high, which means it is higher than $H_{load}$. Also, when the predicted packet loss percentage ($\mathbf{P_L(t+1)}$) is higher than the decided threshold, we can validate the intent. 

Next, intents are executed for multiple combinations of the defined thresholds in $S_{th}$ based on the predicted values of packet loss, load, and power consumption. We calculate the QoS drifts while executing such intents. We form the following look-up table (Table \ref{tab2}):  
\begin{table}[H]

\centering
\caption{Table showing intent types, network state, and QoS drifts.}
\begin{tabular}{|c|c|c|}
\hline
\textbf{Intent types} & \textbf{Network states (thresholds)} & \textbf{QoS drifts} \\ \hline
Energy                         & 0, 1, 1                 & 0                   \\ \hline
Throughput                     & 1, 1, 1                 & 0                   \\ \hline
Energy                         & 1, 1, 1                 & 1                   \\ \hline
Throughput                     & 0, 1, 1                 & 1                   \\ \hline
Throughput                     & 0, 1, 0                 & 0                   \\ \hline
...                            & ...                    & ...                 \\ \hline
Delay                          & 1, 1, 1                 & 0                   \\ \hline
\end{tabular}
\label{tab2}
\end{table}

Based on the lookup table, we either validate or invalidate intents in a supervised manner. We use a threshold selection algorithm (Algorithm 2) to select the thresholds which represent the network states. These thresholds are presented in the second column of the Table \ref{tab2}. 

\begin{algorithm}
\label{algothresh}
\caption{Threshold Selection Algorithm}
\begin{algorithmic}[1]
    \State $H = \{(\text{KPI}_{a_1}, \text{KPI}_{b_1}, M_{th_1}), \dots, (\text{KPI}_{a_n}, \text{KPI}_{b_n}, M_{th_n})\}$
    \State Define $\zeta_a$ and $\zeta_b$ as significance thresholds for detecting significant changes in $\text{KPI}_{a}$ and $\text{KPI}_{b}$, respectively
    
    \State \textbf{Step 1:} Each observation $(\text{KPI}_{a_i}, \text{KPI}_{b_i}, M_{th_i}) \in H$:
    \For{each $i = 1$ to $n$}
        \State Analyze the relationship among: $\text{KPI}_{a_i}, \text{KPI}_{b_i}, M_{th_i}$.
        \State Denote increasing relationship as $R_I$. 
        \State Denote decreasing relationship as $D_I$.
        \State Calculate $\Delta_{RC_a} =  \frac{\text{KPI}_{a_i} - \text{KPI}_{a_{i-1}}}{\text{KPI}_{a_{i-1}}}$. 
        \State Calculate $\Delta_{RC_b} = \frac{\text{KPI}_{b_i} - \text{KPI}_{b_{i-1}}}{\text{KPI}_{b_{i-1}}}$.

        \State \textbf{Condition 1:}
        \If{$R_I = 1$ for $\text{KPI}_{a_i}$ or $\text{KPI}_{b_i}$}
            \State Check whether $\Delta_{RC_a} > \zeta_a$ or $\Delta_{RC_b} > \zeta_b$. 
        \EndIf 

        \State \textbf{Condition 2:}
        \If{$D_I = 1$ for $\text{KPI}_{a_i}$ or $\text{KPI}_{b_i}$}
            \State Check whether $\Delta_{RC_a} < -\zeta_a$ or $\Delta_{RC_b} < -\zeta_b$.
        \EndIf 

        \If{Condition 1 is met}
            \State Set $Th_u = M_{th_i}$ as the upper threshold.
        \EndIf
        \If{Condition 2 is met}
            \State Set $Th_l = M_{th_i}$ as the lower threshold.
        \EndIf
    \EndFor
    
    \State \textbf{Step 2:} Continuously monitor $\text{KPI}_{a_i}$, $\text{KPI}_{b_i}$, and $M_{th_i}$ over time to detect changes
    \State \textbf{Step 3:} Adaptively refine $Th_l$ and $Th_u$
    
\end{algorithmic}
\end{algorithm}

As presented in Algorithm 2, we determine thresholds for a predicted metric by identifying significant changes in its associated \glspl{KPI}. The relationships between the predicted metric and its associated \glspl{KPI} can be classified into two categories: 
\begin{itemize}
    \item Increasing Relationship ($R_I$): A \gls{KPI} has an increasing relationship with the predicted metric if a higher \gls{KPI} value indicates a performance degradation. In such cases, when the {KPI} value rises sharply, it signals potential network issues.
    \item Decreasing Relationship ($D_I$): A \gls{KPI} has a decreasing relationship with the predicted metric if a lower \gls{KPI} value signifies performance degradation. Here, a sudden drop in the \gls{KPI} value signals potential network deterioration. 
\end{itemize}

For example, if we want to select a higher threshold ($H_{load}$) for traffic load, associated crucial metrics with traffic load are delay and packet loss. According to Algorithm 2, $\text{KPI}_a$ in this case is $\textit{delay}$ and $\text{KPI}_b$ in this case is \textit{packet loss}. Since we are seeking a threshold for traffic load, it is denoted as $M_{\text{th}}$. When higher traffic volumes cause a sharp increase in both $\text{KPI}_a$ and $\text{KPI}_b$, it signals potential network performance degradation, particularly in throughput. This point is identified as the high threshold for traffic load, where throughput-enhancing applications can be triggered.

To formally detect significant changes, we compute the \textit{Relative Change Factor} for each \gls{KPI}. For a given \gls{KPI} $x$, its relative change at time step $i$ is defined as:
\begin{equation}
    \Delta_{RC_x} = \frac{\text{KPI}_{x_i} - \text{KPI}_{x_{i-1}}}{\text{KPI}_{x_{i-1}}},
\end{equation}
where $\text{KPI}_{x_i}$ is the current value of KPI $x$, $\text{KPI}_{x_{i-1}}$ is the previous value of KPI $x$, and $\Delta_{RC_x}$ represents the relative change between consecutive time steps.

To determine whether a KPI change is significant, we introduce a thresholding factor, denoted as $\zeta_x$ for each KPI $x$. The value of $\zeta_x$ represents the minimum percentage change required to classify a KPI variation as significant. Specifically, if $\Delta_{RC_x} > \zeta_x$, the KPI has increased significantly. On the contrary, if $\Delta_{RC_x} < -\zeta_x$, the KPI has decreased significantly. The value of $\zeta_x$ is not a fixed constant but is dynamically determined based on historical KPI variations. It is computed via percentile-based thresholding method, where $\zeta_x$ is set based on the historical distribution of KPI fluctuations.

According to Algorithm 2, when a predicted metric (e.g., traffic load) experiences a significant increase in its associated KPIs that have an increasing relationship ($R_I$), it signals potential network performance degradation. In this case, the corresponding value of $M_{\text{th}}$ is marked as the upper threshold ($H_{\text{load}}$). This threshold can be used to trigger throughput-enhancing applications.

Conversely, for KPIs with a decreasing relationship ($D_I$), a significant decrease in their values also signals potential issues. When such a change is detected, the corresponding $M_{\text{th}}$ is marked as the lower threshold $L_{\text{load}}$.

Finally, we present Algorithm \ref{algo3} that summarizes the entire intent validation methodology. 

\begin{algorithm}
\caption{Intent Validation Algorithm}
\label{algo3}
\begin{algorithmic}[1]
    \State \textbf{Input:} Operator’s Prompt: Intent type $i_{t_y}$, Network state $S_N$ (represented by thresholds $N_{th}$). 
    \State \textbf{Output:} Intent validation result (\textbf{Valid} or \textbf{Invalid})

    \State \textbf{Require}: Data extracted from operator’s input and network status $(i'_{t_y}, S'_N)$.
    
    \State \textbf{Step 1:} Identify intent type $i_{t_y}$ and magnitude of increase $\Lambda$ from the operator's input.
    
    \State \textbf{Step 2:} Retrieve network state $S_N$ and check $Q_{drift}$ for each intent type $i_{t_y}$ in the lookup table $T_{\text{lookup}}$.
    
    \For{each entry $(i_{t_y}, S_N, Q_{drift})$ in $T_{\text{lookup}}$}
        \If{$i_{t_y} = i'_{t_y}$ and $S_N = S'_N$}
            \If{$Q_{drift} > 0$}
                \State \textbf{Mark intent as Invalid} 
                \State \textbf{Break} the loop
            \Else
                \State Mark intent as \textbf{Valid}
            \EndIf
        \EndIf
    \EndFor
    
    \State \textbf{Step 3:} If no QoS drift found in any matching entry, return \textbf{Valid}; otherwise, return \textbf{Invalid}.
\end{algorithmic}
\end{algorithm}

\subsection{Application Orchestration Using Hierarchical Decision Transformer}

To this end, we introduced intent processing and validation. Once intents are processed and validated, our proposed scheme extends into realization of the intent through optimizing the network. For this purpose, we introduce \gls{HDTGA}, a hierarchically organized \gls{DT} architecture that pursues goals to achieve instead of returns-to-go in vanilla decision transformers. We aim to initiate and orchestrate network optimizing applications for intent fulfillment in a faster and near-real-time manner at test time. It is possible to use  legacy network optimization intent fulfillment. However, we take this one step further by using DRL-based approaches to fulfill an intent. In real deployments these approaches can co-exist. We have developed the following optimization applications to test the utility of \gls{HDTGA}:
\begin{itemize}
    \item \textbf{Traffic steering:} This application uses a \gls{DQN}-based traffic steering mechanism to optimize performance in multi-\gls{RAT} networks with dual connectivity between LTE and 5G NR \cite{22}. The proposed model dynamically allocates traffic based on \gls{QoS} requirements, specifically targeting throughput and delay.
    \item \textbf{Cell sleeping:} This application enables cell sleeping decisions based on traffic load ratios and the queue length of each \gls{BS} \cite{23}. The objective is to optimize energy efficiency while also ensuring that active \glspl{BS} are not overloaded.  
    \item \textbf{Power allocation:} The power allocation application \cite{25} tries to maximize the total throughput by selecting a power level for each resource block group of each \gls{BS}.
    \item \textbf{Beamforming and power control:} We also employ DQN to develop the beamforming application \cite{26}. It uses \gls{UE} coordinates as states. Actions include selecting beam steering angles and adjusting power levels. The reward function balances throughput and energy efficiency.
    \item \textbf{Energy efficient handover management :} This application \cite{28} is specifically tailored with \gls{DQN} to achieve energy efficiency via optimally tailored handover policies.
\end{itemize}

\begin{figure*}[!t]
\centerline{\includegraphics[width=0.8\linewidth]{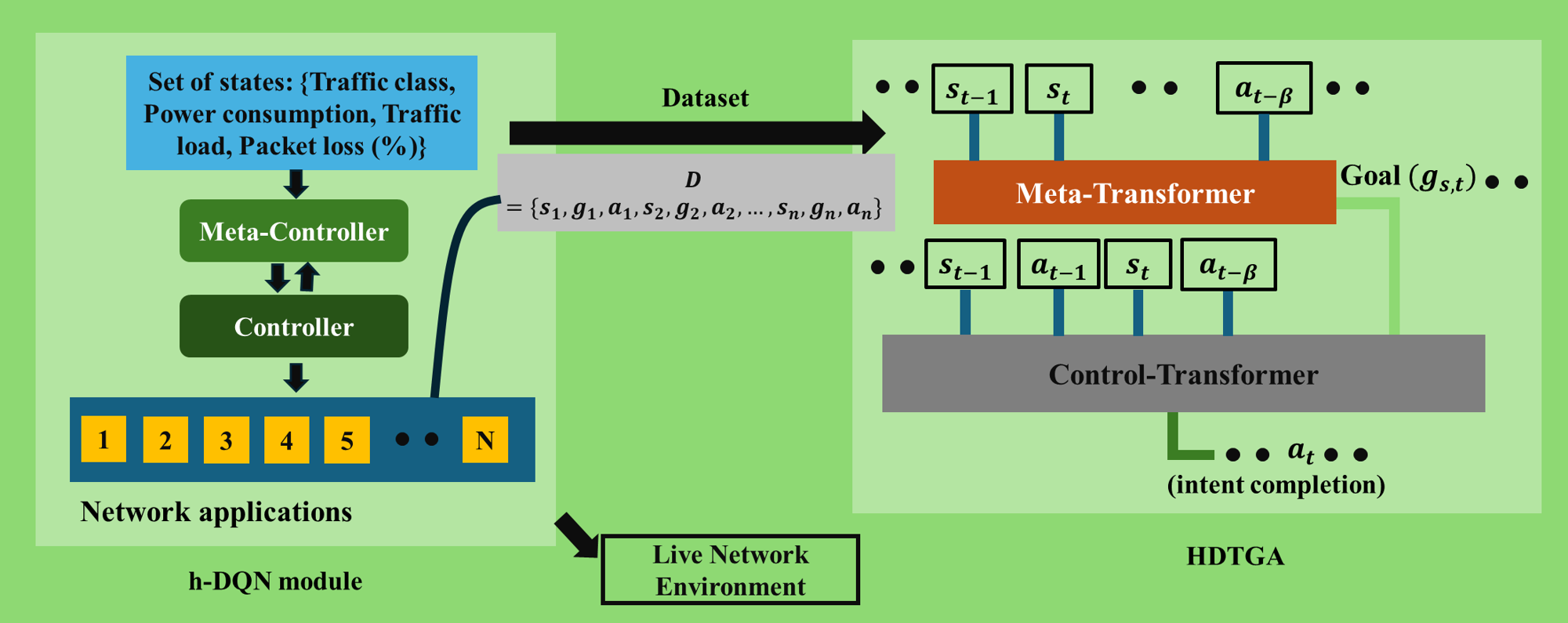}}
\caption{HDTGA architecture with an offline dataset collected from the network environment. }
\vspace{-1.2em}
\label{pta}
\end{figure*}

\begin{algorithm}[!t]
%\begin{small}
\caption{Application Orchestration Using HDTGA}
\label{algo:hDQN_Data_Collection}
\begin{algorithmic}[1]
    \State \textbf{Initialize} dataset $D_{offline}$
    \State \textbf{Initialize} meta-controller and controller in h-DQN framework

    \For{each intent $I \in I$}
        \State Extract goal $g_t \in \mathcal{G}$ from intent $I$ 
        \State Observe current network state $s_t \in \mathcal{S}$
        \State Meta-controller sets $g_t$ and observes $s_t$
        \State $g$ along with $s_t$ is passed to the controller.
        \State Controller chooses $a_t \in \mathcal{A}$ to optimize network
        \State Execute action $a_t$ in the simulated environment
        \State Observe $s_{t+1}$ and receive intrinsic reward $r_{in}$
        \State Record trajectory $tr_i = \{s_t, g_t, a_t, r_{in}\}$
        \State Append $tr_i$ to $D_{offline}$
    \EndFor
    \State \textbf{Return:} $D_{offline}$
    \State \textbf{Require} a language model $M_{\text{LLM}}$ to process custom dataset $D$ with prompts/queries and responses. 
    \State  \textbf{Fine-tune} model: $M_{\text{F-LLM}}$, for responding accurately to network-related queries and intents. 
    \State \textbf {Validtae/ Invalidate}  intent
    \If{Intent is valid}
        \State \textbf{Initialize} $g_t = g_{target}$ from the validated intents
        \State \textbf{Identify} $a_{t-\beta}$ from the offline dataset $D_{offline}$.
        \State \textbf{Use} meta-transformer to predict $a_{t-\beta}$ based on $s_t$, $g_t$
        \[
          a_{t-\beta} = \text{MetaTransformer}(s_{t-1}, s_t, g_{t})
        \]
        \State \textbf{Use} the control-transformer to predict $a_t$ to achieve $g_t$:
        \[
         a_t = \text{ControlTransformer}(s_{t-\omega:t}, g_t, a_{t-\beta})
        \]
        \State Execute action $a_t$
    \EndIf
    \State Initiate and Orchestrate network-optimizing application(s) $A = \{A_{App1}, A_{App2}, \ldots\}$.
\end{algorithmic}
%\end{small}
\end{algorithm}

\glspl{DT} can be extremely efficient since data associated with the past time stamps are also fed as input for taking control actions. However, the need to specify desired returns for each task creates a dependency on external task-specific knowledge, which makes decision transformers less flexible for new environments. It is difficult to generalize across diverse tasks without manually adjusting the desired returns.

The proposed method (HDTGA) eliminates the need to manually specify the returns-to-go. Instead, it uses goals identified from the demonstration dataset to guide the learning process. It has a high-level mechanism that provides the past state ($t-1$), the current state (at $t$), and an important past action that previously reached the goal (partially fulfilling the goal is also acceptable). We use past state sequences to predict the important action every time a new goal is provided, which guides the agent through the task. It provides a way for the low-level transformer to focus on specific actions that are valuable for completing the task. It learns to predict actions that lead to goals, effectively fulfilling an intent from the operator. 

We refer to the transformer on top as the meta-transformer and the transformer on the bottom as the control-transformer. We present Fig. \ref{pta} displaying this bi-level architecture. In the figure, $s_{t-1}$ is the past state, $a_{t-\beta}$ is the useful action in the past that has reached the goal partially or fully, $g_{s,t}$ is the goal to be achieved extracted from the operator intent, and $a_t$ is the action to be taken by the control transformer at time $t$.  

At each time step $t$, the environment provides a state $s_t \in \mathcal{S}$, and the aim is to learn a policy $\pi(a_t | s_t, g_t)$ that guides towards a goal $g_t \in \mathcal{G}$. Each component, states $s_t$, goals $g_t$, and important past actions $a_{t-\beta}$ is embedded and fed into the transformer. Here, $\beta$ is the time offset of the action from the past that partially achieved a goal. Let $E(\cdot)$ denote the embedding function and $P(\cdot)$ the positional encoding:

\begin{itemize}
    \item State embeddings: $E(s_t) + P(t)$
    \item Goal embeddings: $E(g_t) + P(t)$
    \item Important action embeddings: $E(a_{t-\beta}) + P(t-\beta)$
\end{itemize}

Thus, the transformer input at time $t$ is:
\begin{equation}
\text{tr}_{i_t} = \left[ E(s_t) + P(t), \, E(g_t) + P(t), \, E(a_{t-\beta}) + P(t-\beta) \right]
\end{equation}

The self-attention mechanism within the transformer computes attention weights across these embeddings, allowing the model to attend to important past actions effectively.

The meta-transformer policy $\pi_{\phi}^{meta}$ identifies a significant past action $a_{t-\beta}$ which previously achieved the goal partially or fully, based on a series of past states $s_{t-\omega:t-1}$, the current state ($s_t$), and the desired goal ($g_t$). 
\begin{equation}
    a_{t-\beta} = \pi_{\phi}^{meta} (s_{t-\omega:t-1},s_t,g_t),
\end{equation}
where $\omega$ is the window length of past states considered.

The control-transformer, responsible for predicting actions at each time step $t$, takes in the sequence of encoded states, goals, and important past actions. The predicted action $a_t$ is derived by conditioning on past states and the goal:
\begin{equation}
    a_t = \pi_{\theta}^{control}(s_{t-K:\omega}, g_t, a_{t-\beta})
\end{equation}
The predicted action ($a_t$) in this case is the action of selecting non-conflicting applications for initiation to fulfill an intent. We propose Theorem \ref{th:HDTGA} associated with the optimality of the proposed \gls{HDTGA}. 

\begin{theorem} \label {th:HDTGA}
The Hierarchical Decision Transformer with Goal Awareness (HDTGA) learns a policy $\pi_{\theta}(a_t | s_{\leq t}, g_t)$ such that the expected return $V^{\pi_{\theta}}(s_0, g_0)$ is within $\epsilon$ of the optimal expected return $V^{*}(s_0, g_0)$ for all initial states $s_0$ and goals $g_0$, i.e.,
\[
| V^{\pi_{\theta}}(s_0, g_0) - V^{*}(s_0, g_0) | \leq \epsilon,
\]
provided that the model capacity and training procedure are sufficient to minimize the empirical loss to an acceptable level.
\end{theorem}

A proof of Theorem \ref{th:HDTGA} is provided in Appendix A. We summarize the entire proposed method putting special focus on the application orchestration part in Algorithm 4.

\section{Performance Evaluation}

This section presents a detailed performance evaluation of the proposed method. It first describes the simulation setup. Then, it provides results to show the effectiveness of the three-step methodology: intent processing, intent validation, and application orchestration for intent execution.

\label{s5}
\subsection{Simulation Setup}
The simulation setup in this study consists of a macro cell surrounded by densely deployed small cells in a multi-\gls{RAT} environment. The cells serve a total of 60 users. Configurations associated with the 5G NR and LTE \glspl{RAT} are presented in Table \ref{sim}. 

We consider four traffic types: video, gaming, voice, and an Ultra-Reliable Low-Latency Communication (URLLC)-based use case (vehicle-to-\gls{BS} data traffic). Each of these traffic types is characterized by specific inter-arrival times for packets: 12.5 ms, 40 ms, 20 ms, and 0.5 ms, respectively \cite{29}. The arrival of data packets follows different distributions: Pareto for video \cite{29}, Uniform for gaming \cite{29}, and Poisson for both voice \cite{46} and the URLLC \cite{47} use case. In our system, we adopt different antenna configurations and frequency bands to optimize the performance across mid-band (3.5 GHz) and high-band (30 GHz) scenarios. For the mid-band, a Uniform Linear Array (ULA) with 64 antennas is deployed to enhance coverage and maintain high spectral efficiency, given the relatively lower propagation losses at this frequency. On the contrary, a Uniform Planar Array (UPA) with 128 antennas is utilized for advanced beamforming capabilities for the high-band. The bandwidths considered for these configurations are 60 MHz for the mid-band and 100 MHz for the high-band. Table \ref{sim} further summarizes these simulation settings along with the necessary parameters of the algorithms used in this research. 

\begin{table}[!t]
    \centering
    \caption{Simulation and Algorithmic hyperparameter Settings}
    \begin{tabular}{|l|l|}
         \hline
         \textbf{\underline{5G NR}} & \\
         Bandwidth & $50$ and $100$ MHz \\
         Carrier frequency &  $3.5$ and $30$ GHz \cite{27} \\
         Max transmission power & $43$ dBm \cite{20}\\
         Subcarrier spacing & $15$ and $60$ KHz \cite{30} \\
         \hline
         \textbf{\underline{LTE}} & \\
         Bandwidth & $40$ MHz \\
         Carrier frequency & $800$ MHz \\
         Max transmission power & $38$ dBm \cite{20}\\
         Subcarrier spacing & $15$ kHz \\
         \hline
         \textbf{\underline{\gls{DRL} parameters (Network Applications)}} & \\
         Batch size, Initial exploring steps & $32$, $3000$ \\
         Learning rate ($\alpha$), discount factor ($\gamma$) & $0.5$, $0.9$ \\
         \hline
         \textbf{\underline{Informer (Prediction)}} & \\
         Learning rate $(\alpha)$ & $10^{-4}$ \cite{11} \\
         Batch size & $32$ \cite{11}\\
         Encoder and decoder layers & $4$ and $2$ \cite{11}\\
         Multi-head attention in encoder & $16$ heads \cite{11}\\
         Multi-head attention in decoder & $8$ heads \cite{11}\\
         Dropout probability learning rate & $0.1$ and $1e^{-4}$ \cite{11}\\
         \textbf{\underline{HDTGA (Common)}} & \\
         Batch size & $32$\\
         Q-value alignment weight & $0.5$ \\
         \textbf{\underline{HDTGA (Meta-transformer)}} & \\
         Number of attention heads and layers & $8$ and $4$\\
         Dropout and learning rate & $0.1$ and $1 \times 10^-4$\\
         \textbf{\underline{HDTGA (Control-transformer)}} & \\
         Number of attention heads and layers & $8$ and $4$\\
         Dropout and learning rate & $0.1$ and $1 \times 10^-4$\\
         \hline
    \end{tabular}
    \label{sim}
\end{table}

\subsection{Simulation results}

\subsubsection{Intent and Query Processing}

\begin{figure}[htp]
\centerline{\includegraphics[width=0.8\linewidth]{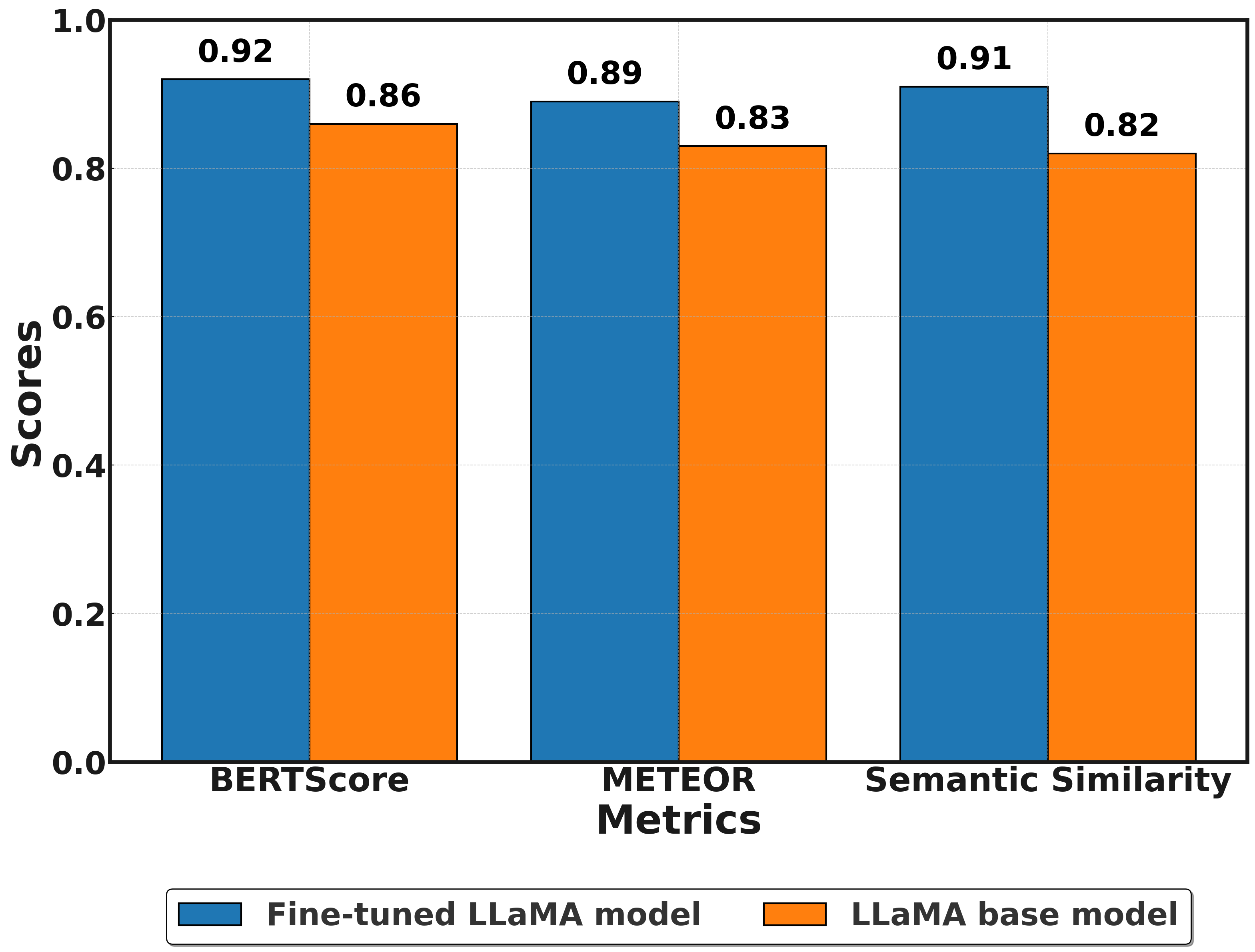}}
\caption{Performance comparison among the proposed intent and query processing methodology and the LLM baseline.}
\vspace{-1.2em}
\label{meteor}
\end{figure}

The first set of results in this work is associated with the LLM-fine tuning based on QLoRA. The LlaMa base model is used as a baseline in this case \cite{21}. Three crucial metrics have been used for performance comparison. These are \textit{BERTScore} \cite{BERTScore}, \textit{METEOR} \cite{Meteor}, and \textit{semantic similarity} \cite{SS}. These are commonly used metrics for evaluating LLM performance, especially in tasks like text generation, summarization, machine translation, and natural language understanding.

To evaluate the model performance in terms of \textit{BERTScore}, we start by taking the LlaMA-generated output and a reference text (human-annotated correct response). After that, a pre-trained BERT model is used to generate embeddings for the words in both the generated output and reference text. Lastly, we calculate the token-level cosine similarity between corresponding embeddings, following the standard \textit{BERTScore} process. The BERTScores for the LlaMa base model and the QLoRA-based fine-tuned model reflect how well the LlaMA output aligns semantically with the reference text. Since BERTScore captures fine-grained token-level semantics, it provides insights into whether LlaMA’s generated text maintains the same meaning as the reference. As it can be observed from Fig. \ref{meteor}, a BERTScore of 0.92 has been achieved by the fine-tuned model compared to the base model's 0.86. This increased score indicates that the fine-tuned LlaMA model is capturing the intended meaning of the reference text more accurately than the base model.

\begin{figure*}[!t]
\centerline{\includegraphics[width=1\linewidth]{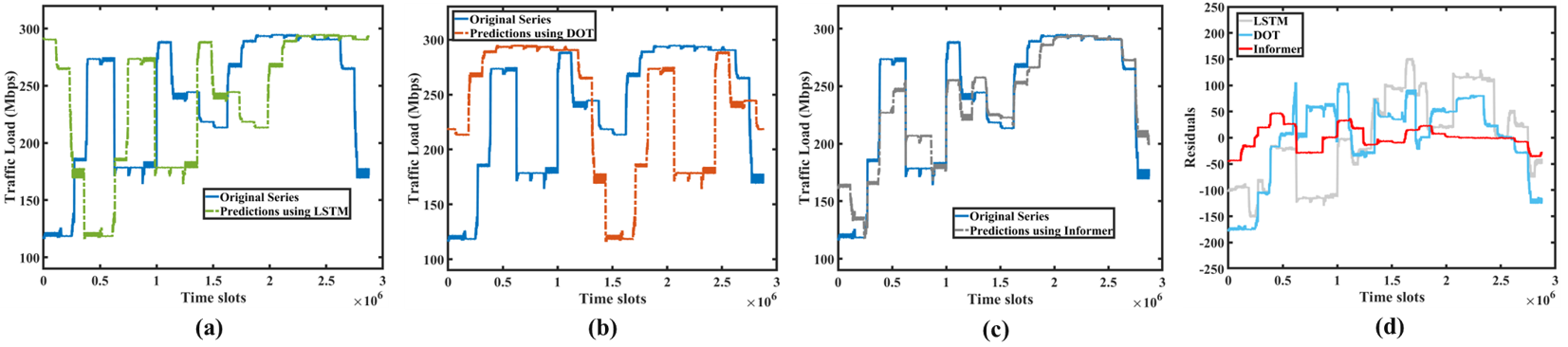}}
\vspace{-1.2em}
\caption{Prediction curves: (a) LSTM, (b) DOT, and (c) Informer and residual plot: (d).}
\label{tl-timeseries}
\end{figure*}

\begin{figure*}[!t]
\centerline{\includegraphics[width=0.75\linewidth]{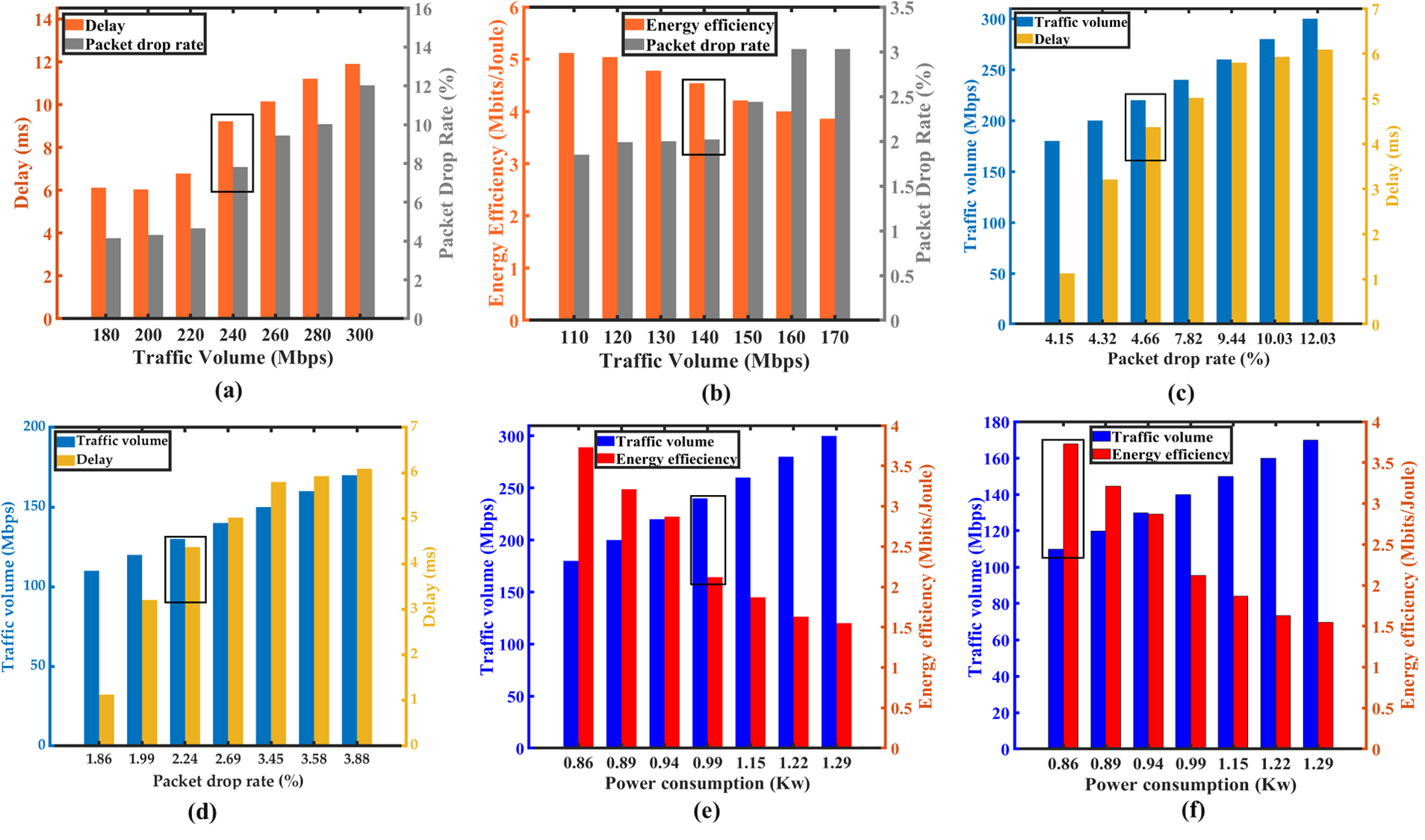}}
\vspace{-1.2em}
\caption{Threshold selection via performance observation: (a) $H_{load}$, (b) $L_{load}$, (c) $H_{loss}$, (d) $L_{loss}$, (e) $H_{pc}$, and (f) $L_{pc}$ }
\label{fullth}
\end{figure*}

The \textit{METEOR} score of 0.89 for the QLoRA-based fine-tuned LlaMA model compared to the base LlaMA model’s score of 0.83, indicates a clear improvement in the semantic alignment of the generated outputs after fine-tuning (See Fig. \ref{meteor}). This increase from 0.86 to 0.89 suggests that the fine-tuned LlaMA model is generating outputs that better match the reference text in terms of meaning, lexical choice (including synonyms), and word order. For intent-driven management, such as application orchestration in RAN, it is important that the model accurately conveys specific terms and instructions. The higher \textit{METEOR} score for the fine-tuned model shows that it is better at preserving and incorporating these critical terms.

The third metric, \textit{semantic similarity} measures how similar two pieces of text are in meaning, rather than exact wording. From Fig. \ref{meteor}, we can see that the proposed intent processing method outperforms the baseline in terms of semantic similarity too. A higher semantic similarity score means the fine-tuned model can better recognize and maintain the specific intent, even when the phrasing or structure might vary slightly. This is critical because misunderstandings or deviations from the intended instructions can lead to unwanted configuration of applications. 

\subsubsection{Intent validation}

The core of the intent validation process in our proposed methodology is the use of a transformer architecture (Informer) to predict traffic load, power consumption, and packet loss percentage. To justify our choice of an advanced time-series forecasting method like the Informer, we compare the performance of the Informer with \gls{LSTM} \cite{33} and Generative Pre-trained Transformer 2 (GPT2)'s Decoder Only Transformer (DOT) \cite{34}. Fig. \ref{tl-timeseries}a, b, and c present the prediction curve of Informer, GPT2-DOT, and LSTM respectively against the actual traffic data. Here, prediction results are presented for traffic load only. Similar graphs and performance can also be obtained for the other two parameters: power consumption and packet loss.  

Next, we use the time series residual plot to evaluate the performance of the Informer architecture. Residuals are computed by subtracting the predicted values from the actual observations in the time series. When plotted over time, a fluctuating pattern (such as increasing or decreasing residuals) indicates that the model has not fully captured certain temporal structures in the data. As shown in Fig. \ref{tl-timeseries}d, the residuals from the proposed Informer-based wireless traffic prediction remain dispersed around zero, showing the model's effectiveness in capturing the underlying patterns. On the other hand, residuals of the baselines fluctuate highly. This suggests that Informer's predictions are closer to the actual observed values compared to the baselines. Next, the Mean Absolute Error (MAE) metric is used to compare the prediction accuracy of time-series models. MAE is calculated by measuring the average absolute difference between observed and predicted values. The MAE for LSTM, DOT, and Informer is 70.5, 64.9, and 16.1, respectively. Informer achieves the lowest MAE, which indicates superior predictive accuracy. Similarly, Informer outperforms GPT2’s DOT and LSTM in predicting power consumption and packet loss (\%). 

\begin{figure*}[!t]
\centerline{\includegraphics[width=0.65\linewidth]{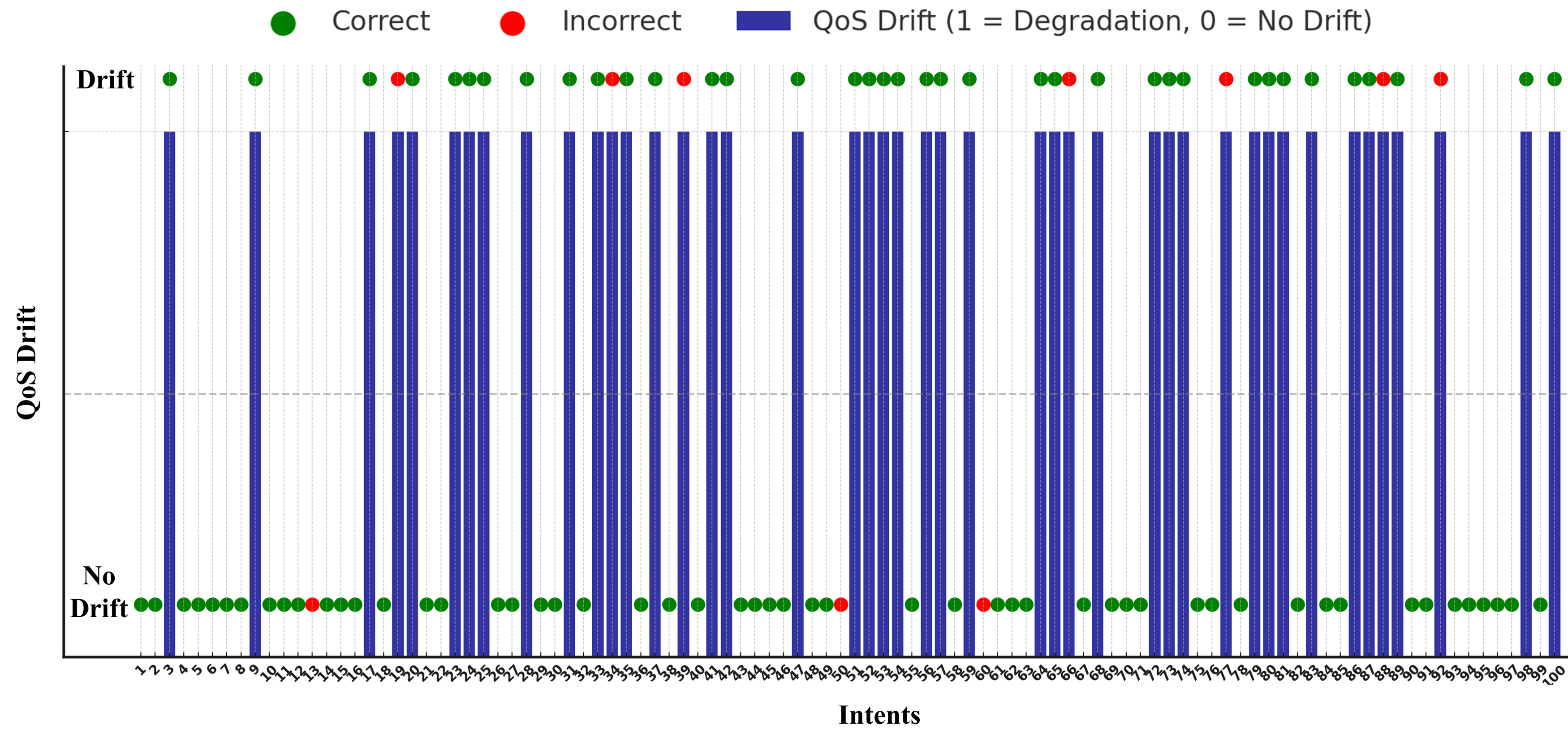}}
\vspace{-1.2em}
\caption{Validation of intents based on QoS drift calculation.}
\vspace{-1.2em}
\label{IV_acc}
\end{figure*}

\begin{figure*}[!t]
\centerline{\includegraphics[width=0.75\linewidth]{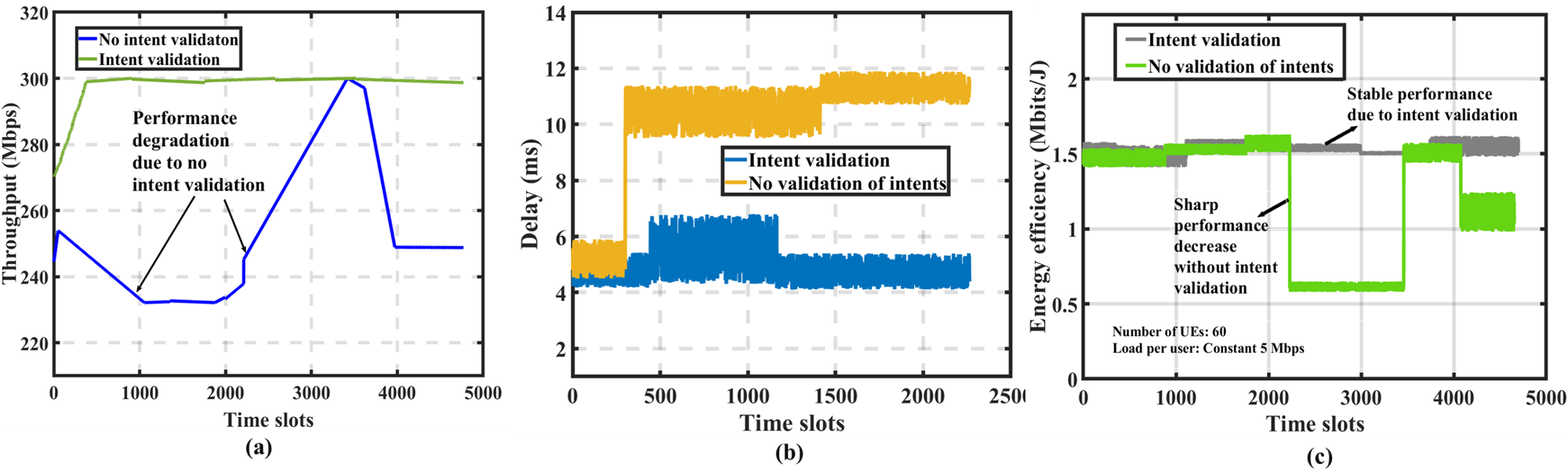}}
\vspace{-1.2em}
\caption{Impact of intent validation: (a) Throughput, (b) Delay, and (c) Energy efficiency}
\label{IV}
\end{figure*}

In summary, Informer outperforms two prominent benchmark models in the literature, excelling in both prediction accuracy and network efficiency. LSTM struggles to capture long-term dependencies effectively due to performance degradation with extended sequences. Even with extensive hyperparameter tuning, LSTM could not match Informer’s efficiency on long sequences. Transformer models, on the other hand, utilize their self-attention mechanism to handle long-range dependencies more proficiently. Our approach uses Informer, an encoder-decoder Transformer model. It uses ProbSparse self-attention and a generative style decoder. With explicit focus on capturing both short-term and long-term temporal dependencies through mechanisms like sequence distillation and multi-scale modeling, Informer provides way better performance than GPT2's DOT.

Predictions are categorized as high or low based on selected thresholds. Algorithm 2 determines these thresholds by detecting sudden changes in key metrics. Fig. \ref{fullth} presents bar plots for setting thresholds on traffic load (\ref{fullth}a, \ref{fullth}b), power consumption (\ref{fullth}c, \ref{fullth}d), and packet drop rate (\ref{fullth}e, \ref{fullth}f). Rectangular boxes indicate the chosen threshold values.

Next, Fig. \ref{IV_acc} shows the accuracy of the intent validation method. The X-axis represents a subset of intents, while the Y-axis shows corresponding QoS drifts. Red circles indicate incorrect validations, where intents were accepted despite QoS drifts. Only 12 out of 100 intents were incorrectly validated, achieving 88\% accuracy.

Without intent validation, misaligned intents impact \glspl{KPI}. Fig. \ref{IV} illustrates fluctuating throughput (\ref{IV}a), delay (\ref{IV}b), and energy efficiency (\ref{IV}c) without intent validation. With intent validation, KPI curves are smoother, ensuring stable performance.

\subsubsection{Application Orchestration}

\begin{figure*}[!t]
\centerline{\includegraphics[width=1\linewidth]{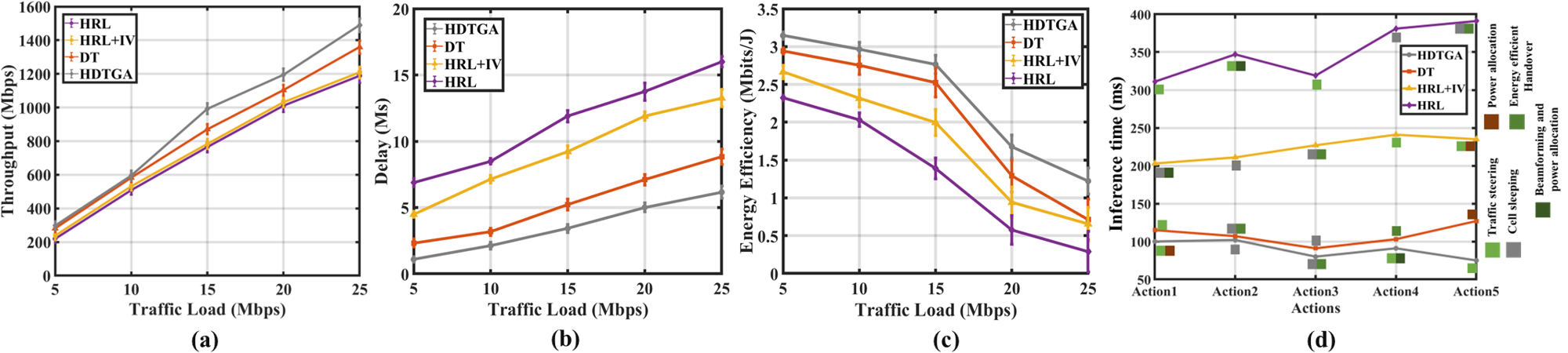}}
\vspace{-1.2em}
\caption{Performance comparison of the proposed application orchestration methodology  against the baselines in terms of (a) Throughput, (b) Delay, and (c) Energy efficiency) (d) Action inference time.}
\label{KPI}
\end{figure*}

We compare the application orchestration performance of the proposed HDTGA with the baseline algorithms. The first baseline algorithm is using HRL that has no intent validation \cite{41}. The second baseline scheme is using the same HRL algorithm but has an intent validation scheme. Please refer to Appendix B for more details on the HRL algorithm used as a baseline. Lastly, the third baseline algorithm used to compare the performance of HDTGA in terms of application initiation and orchestration is the vanilla DT \cite{12}. We consider three crucial KPIs: throughput, delay, and energy efficiency. 

Fig. \ref{KPI}a presents the performance comparison of the proposed HDTGA with the baseline schemes in terms of throughput. It gains $9.5\%$, $23.2\%$, and $25.2\%$ increase compared to the DT, HRL with intent validation, and HRL without intent validation, respectively. Furthermore, the proposed method achieves better performance in terms of delay as it achieves $30.4\%$, $53.5\%$, and $61.5\%$ reduction in delay compared to the same baselines as presented in Fig. \ref{KPI}b.  Lastly, we observe the performance of the proposed method in terms of energy efficiency. An increased energy efficiency of $41.1\%$, $46.72\%$, and $77.4\%$ is observed compared to baseline algorithms. HDTGA’s meta-transformer identifies significant past actions that previously achieved similar goals or sub-goals. Identifying the goal relevant actions eliminates unnecessary steps minimizing time spent on suboptimal tasks. As a result, the system experiences reduced delay, improved throughput, and energy efficiency. Furthermore, the proposed hierarchical structure limits action prediction to relevant historical actions rather than a full action-space search. It is highly necessary for delay-sensitive applications. This also results in less resource consumption per decision, enhancing energy efficiency. Lastly, Fig. \ref{KPI}d presents a plot where we can see that the action inference time for the proposed method is lower than the baselines. This shows that the HDTGA is more computationally efficient. It enables faster decision-making, which is critical for near-real-time network automation tasks.

Multiple different trajectories (state, goal, action trios) may reach the goal partially. In such cases, we can rank the goal achievement. An exponential deviation function is used to calculate the goal deviation. 
\begin{equation}\label{gd}
    G_d = s_f \cdot \left( e^{\left| \frac{G_{f} - G_{a}}{G_{f}} \right|} - 1 \right),
\end{equation}
where $G_f$ is the fixed goal provided by the operator to achieve, and $G_a$ is the achieved goal, $s_f$ is a scaling factor that controls the growth rate of the deviation. For example, if the goal to achieve based on the intent from an operator is $7$ ms in terms of delay, the application orchestration algorithm can take action to reach such a goal. However, there is no guarantee that the system will be able to achieve that goal exactly. (\ref{gd}) calculates the deviation from the goal that was intended to be achieved.   

\begin{figure}[!t]
\centerline{\includegraphics[width=0.60\linewidth]{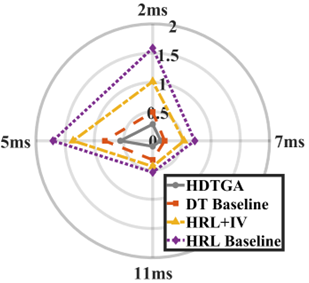}}
\vspace{-1.2em}
\caption{Performance comparison in terms of goal (delay goals) deviations: proposed method vs. the baselines.}
\vspace{-1.2em}
\label{ga}
\end{figure}

In Fig. \ref{ga}, we plot the goal deviations for the proposed HDTGA and the baseline algorithms via a spider plot. The goals are associated with delays (2, 5, 7, and 11 ms). It can be observed that the proposed method results in the least amount of goal deviation. As the delay target is relaxed (e.g.,7ms to 11ms), all methods naturally show improved success in meeting the performance criteria due to increased flexibility. However, the proposed HDTGA outperforms the baselines, showing its effectiveness even under stricter delay constraints. This suggests that HDTGA is better suited for scenarios requiring stringent latency. We observed a similar performance achievement of HDGTA in terms of the other crucial KPIs such as throughput and energy efficiency too. This suggests that HDTGA maintains better alignment with the operator-defined goals, showing higher precision in achieving specific network targets. 

\section{Conclusions}
\label{s6}
This work proposes an intent-driven network management framework utilizing GenAI algorithms to address the complexities of 5G and future 6G networks. Our approach translates human-operator intents into actionable optimization policies, resulting in efficient intent processing, validation, and network policy generation. First, a QLoRA-based fine-tuned LLM enables efficient and accurate intent processing in resource-constrained environments. Second, an Informer-based predictive model anticipates three key parameters: power consumption, traffic load, and packet drop rate to enable real-time intent validation. Finally, the HDTGA framework optimizes application selection and orchestration that results in significant improvement of the network performance.
Experimental results show clear advantages of the proposed scheme: BERTScore and semantic similarity improved by 6\% and 9\%, while predictive intent validation achieved 88\% accuracy in eliminating performance-degrading intents. Additionally, HDTGA-based policy generation increased throughput by at least 19.3\%, reduced delay by 48.5\%, and improved energy efficiency by 54.9\% over the baselines. These findings highlight our framework’s potential to enable adaptive, on-demand service deployment in dynamic, resource-intensive 6G environments.

\section*{Acknowledgement}
This work has been supported by MITACS and Ericsson, NSERC Canada Research Chairs program and NSERC Collaborative Research and Training Experience Program (CREATE) under Grant 497981.

\bibliographystyle{IEEEtran}
\bibliography{reference.bib}

\clearpage
\section*{\textbf{Appendix A}}
\label{B}

\textit{Proof.} Consider a Markov Decision Process (MDP) defined by the tuple $(\mathcal{S}, \mathcal{A}, \mathcal{G}, P, R, \gamma)$, where:
\begin{itemize}
    \item $\mathcal{S}$ is a finite set of states,
    \item $\mathcal{A}$ is a finite set of actions,
    \item $\mathcal{G}$ is a set of goals,
    \item $P: \mathcal{S} \times \mathcal{A} \times \mathcal{S} \rightarrow [0,1]$ is the state transition probability,
    \item $R: \mathcal{S} \times \mathcal{A} \times \mathcal{G} \rightarrow \mathbb{R}$ is the reward function dependent on goals,
    \item $\gamma \in [0,1)$ is the discount factor.
\end{itemize}
Assume that:
\begin{enumerate}
    \item The reward function $R(s, a, g)$ is bounded: $|R(s, a, g)| \leq R_{\text{max}}$.
    \item The state transition probabilities and reward function satisfy the Lipschitz condition with respect to a suitable metric.
    \item The offline dataset $\mathcal{D}_{\text{offline}}$ covers the state-action-goal space sufficiently, ensuring that all state-action-goal triples have been observed with non-zero probability.
\end{enumerate}

We will prove the theorem by leveraging the properties of sequence modeling in the HDTGA architecture and its relation to the Bellman optimality equation.

The optimal action-value function $Q^{*}(s, a, g)$ satisfies the Bellman equation:
\[
Q^{*}(s, a, g) = \mathbb{E}_{s'} \left[ R(s, a, g) + \gamma \max_{a'} Q^{*}(s', a', g) \right],
\]
where $s'$ is the next state resulting from action $a$ in state $s$.

The HDTGA policy $\pi_{\theta}$ is designed to predict actions based on the sequence of past states $s_{\leq t}$, goals $g_t$, and important past actions $a_{t - \beta}$ identified by the meta-transformer:
\[
a_t = \pi_{\theta}(s_{\leq t}, g_t, a_{t - \beta}).
\]
The meta-transformer helps focus the policy on relevant past experiences that are critical for achieving the goal $g_t$.

Assuming that the transformer model has sufficient capacity, it can approximate any function, including the optimal policy $\pi^{*}$. Specifically, for any $\Gamma > 0$, there exists a parameter $\theta$ such that:
\[
\mathbb{E}_{(s, g) \sim \mathcal{D}} \left[ \| \pi_{\theta}(s_{\leq t}, g_t, a_{t - \beta}) - \pi^{*}(s, g) \|^2 \right] \leq \Gamma,
\]
where $\mathcal{D}$ is the distribution of states and goals in the offline dataset.

The error in the policy translates to an error in the value function $V^{\pi_{\theta}}(s, g)$. Using standard results from approximate dynamic programming, the difference between the value functions satisfies:
\[
| V^{\pi_{\theta}}(s, g) - V^{*}(s, g) | \leq \frac{2 \Gamma R_{\text{max}}}{(1 - \gamma)^2}.
\]
This inequality arises from the Lipschitz continuity of the Bellman operator and the contraction property with factor $\gamma$.

The assumption that the offline dataset $\mathcal{D}_{\text{offline}}$ sufficiently covers the state-action-goal space ensures that the model can learn the necessary state-action dependencies to approximate $\pi^{*}$. Specifically, for all $(s, a, g)$, the dataset provides samples such that:
\[
P_{\mathcal{D}}(s, a, g) > 0.
\]

Combining the above results, we have:
\[
| V^{\pi_{\theta}}(s, g) - V^{*}(s, g) | \leq \epsilon = \frac{2 \delta R_{\text{max}}}{(1 - \gamma)^2}.
\]
By choosing $\Gamma$ appropriately (through model capacity and training), we can make $\epsilon$ arbitrarily small.

\section*{\textbf{Appendix B}}
\label{D}

We use FeUDal Network as our HRL baseline to compare the application orchestration performance of the proposed HDTGA algorithm. Structure of FeUdal Network for our use case is as follows: 

\textbf{Perceptual Module $(f_{\text{percept}})$:} It processes network observations $(x_t)$ into a shared latent representation $(z_t)$. It takes traffic classes, load, power consumption, packet drop percentage as input and outputs $(z_t)$. 

\textbf{Manager:} Operates at a lower temporal resolution, setting directional goals for the Worker. It learns high-level latent goals $(g_t)$ to improve network performance. Updates take place via transition policy gradients:
\[
\nabla g_t = A_t \cdot \nabla_\theta \cos(\vec{s}_{t+c} - \vec{s}_t, g_t(\theta))
\]

\textbf{Worker:} It executes primitive actions to achieve goals set by the Manager. Takes latent state $(z_t)$, current observation $x_t$, and Manager’s goal $g_t$ as input outputting probabilities over primitive actions optimized by intrinsic rewards: 
\[
R_I = \frac{1}{c} \sum_{i=1}^{c} \cos(\vec{s}_t - \vec{s}_{t-i}, g_{t-i})
\]

The process of implementing a FeUdal Network for HRL is as follows: 
\begin{itemize}
    \item Manager observes the state $s_t$
    \item Predicts a directional goal ($g_t$) in latent space.
    \item Goal communicated to the Worker.
    \item Worker receives $g_t$ , current state, and observation.
    \item Selects (worker) an action $a_t$ using its policy.
    \item Executes action in the environment.
    \item Environment provides extrinsic reward ($R_E$).
    \item Intrinsic reward $R_I$ assigned to the Worker for achieving directional goals.
    \item Manager updates goal-generation policy using transition policy gradients.
    \item Worker updates action-selection policy using advantage actor-critic.
\end{itemize}

Our goals are already defined abstractly (e.g., ``Increase throughput by 10\%," ``Minimize delay to 1 ms"). These goals can be numerically represented as directional changes in specific performance metrics:
$$
g_t=\{\Delta \text{Throughput}, \Delta \text{Energy efficiency}, \ldots ,\Delta \text{Delay}\}
$$
For example: 
\begin{itemize}
    \item \textbf{Intent:} ``Increase throughput by $10\%$."
    \item \textbf{Current throughput:} $200$ Mbps.
    \item \textbf{Goal vector:}$g_t = \{ 220 \text{ Mbps}, 0, 0 \}$ (indicating only throughput improvement).
\end{itemize}

To embed goals into latent space a learnable linear transformation ($\phi$) is used to map goals into a latent embedding:
$$
w(t) = \phi(g_t) = Wg_t.
$$
Here, $W$ is a learned weight matrix. This ensures that abstract goals are translated into a compact, meaningful representation.

The Worker uses the embedded goal $w_t$ to decide which primitive actions (e.g., traffic steering, beamforming) to execute. The goal embedding is multiplied with the action embeddings $U_t$:
$$
\pi(a_t | s_t, g_t) = \text{Softmax}(U_t w_t)
$$

The embedding $\phi$ and worker’s action embedding $U_t$ are trained together.

\clearpage
\onecolumn
\printglossary[type=\acronymtype,style=long]
\clearpage
\twocolumn

\end{document}